\documentclass[aoas]{imsart}

\RequirePackage{amsthm,amsmath,amsfonts,amssymb,float}
\RequirePackage[authoryear]{natbib}
\RequirePackage[colorlinks,citecolor=blue,urlcolor=blue]{hyperref}
\usepackage{graphicx}
\RequirePackage{graphicx}
\usepackage{algorithm}
\usepackage{algpseudocode}
\usepackage{enumitem} 

\startlocaldefs
\theoremstyle{plain}

\newtheorem{theorem}{Theorem}[section]

\newtheorem{assump}{Assumption}

\theoremstyle{remark}

\endlocaldefs

\begin{document}

\begin{frontmatter}
\title{Targeted Maximum Likelihood Estimation for Integral Projection Models in Population Ecology}
\runtitle{}

\begin{aug}
\author[A]{\fnms{Yunzhe}~\snm{Zhou}\ead[label=e1]{ztzyz615@berkeley.edu}} \and
\author[B]{\fnms{Giles}~\snm{Hooker}\ead[label=e2]{ghooker@wharton.upenn.edu}\orcid{0000-0003-2648-1167}}

\address[A]{Department of Biostatistics,
University of California, Berkeley\printead[presep={,\ }]{e1}}

\address[B]{Department of Statistics and Data Science,
University of Pennsylvania\printead[presep={,\ }]{e2}}
\end{aug}

\begin{abstract}
Integral projection models (IPMs) are widely used to study population growth and the dynamics of demographic structure (e.g. age and size distributions) within a population.These models use data on individuals' growth, survival, and reproduction to predict changes in the population from one time point to the next and use these in turn to ask about long-term growth rates, the sensitivity of that growth rate to environmental factors, and aspects of the long term population such as how much reproduction concentrates in a few individuals; these quantities are not directly measurable from data and must be inferred from the model.  Building IPMs requires us to develop models for individual fates over the next time step -- Did they survive? How much did they grow or shrink? Did they Reproduce? --  conditional on their initial state as well as on environmental covariates in a manner that accounts for the unobservable quantities that are are ultimately interested in estimating. 

Targeted maximum likelihood estimation (TMLE) methods are particularly well-suited to a framework in which we are largely interested in the consequences  of models.  These build machine learning-based models that estimate the  the probability distribution of the data we observe and define a {\em target of inference} as a function of these. The intial estimate for the distribution is then modified by tilting in the direction of the efficient influence function to both de-bias the parameter estimate and provide more accurate inference. In this paper, we employ TMLE to develop robust and efficient estimators for properties derived from a fitted IPM, with a particular focus on long-term stable population growth, its elasticity to fecundity, and the expected growth rate under year-specific covariates. Mathematically, we derive the efficient influence function and formulate the paths for the least favorable sub-models. Empirically, we conduct extensive simulations and demonstrate its efficacy using real data from both long term studies of Idaho steppe plant communities and experimental Rotifer populations.
\end{abstract}


\end{frontmatter}

\section{Introduction}

Integral projection models (IPMs) describe the changes over discrete time in a population that is structured by a continuous state variable at the individual level \citep{easterling2000size,ellner2006integral,ellner2007stochastic,gelfand2013scaling, 2016:ellne.child.rees}. 
IPMs  are a powerful tool that translate regression models for individual vital rates, such as survival, growth, and fecundity into long-term characteristics of the population. IPMs play an important role in ecology by offering a mechanistic approach that allows for a comprehensive understanding and connection of biological processes across different scales \citep{merow2014advancing}. A key challenge here is that the things we want to know about involve time scales that are much longer than the data we have to hand; they can only be estimated as consequences of models built on short term data and we have no means of externally validating their conclusions. This makes it critical that both modeling and uncertainty quantification are done well.  Here we adapt the tools of targeted maximum likelihood from causal inference as a natural framework in which to develop inference for conclusions derived from IPMs.

The process of constructing IPMs usually starts with obtaining longitudinal data describing individuals’ vital rates: how much do they survive, grow and reproduce from one time point to the next?  The core of an IPM is the kernel function that explains how an individual's state at one point influences its own state and that of its offspring at a later time. It usually consists of two components: a survival/growth kernel and fecundity kernel. The first of these quantifies the likelihood of an individual surviving throughout the census period and, if so, the probability distribution of its size at the next time point. The second represents the number of offspring generated by reproductive individuals during the census period, along with the size distribution of these new offspring. The  functions  making  up the  kernel  can  often  be  estimated  by  regression that relates the state of an individual along with additional covariates to its vital rates \citep{metcalf2003evolutionary}. Matrix projection models (MPMs) are an earlier methodology \citep{caswell2001matrix} that assumes the individuals are in discrete stages. Mathematically, MPMs and IPMs are similar objects since the numerical integral method is usually used to discretize the kernel of IPM when the kernel functions are not analytically integrable \citep{easterling2000size}. The main difference between them is that IPMs are usually parameterized by regression models, while MPM typically estimates the probabilities directly from observed transitions. 

IPMs require the estimation of regression models. The assumptions underlying the chosen parametric regressions must be carefully evaluated, as they can introduce biases if not properly validated against empirical data. Traditional methods use specific evaluation metrics, such as cross-validation, to select the model that best optimizes these criteria. While cross-validation improves the prediction performance of each model component, it does not specifically focus on the target of inference. Model selection is inherently purpose-driven, with different criteria being optimal for different purposes. For example, the optimal model complexity for estimating a response, such as the long-term growth rate of a species, differs from that needed for estimating the gradient of that response, e.g. the sensitivity of the long-term  growth rate to minor changes in fecundity. These distinctions necessitate exploring suitable approaches that focus the estimation process on the target of inference.

Targeted Maximum Likelihood Estimation (TMLE) is a statistical estimation method designed to improve the estimation of model consequences within complex and high-dimensional data structures. A key component of TMLE is defining the target of inference as a consequence of the data distribution, rather than of internal components of a model \citep{van2006targeted}. In that sense IPMs, in which data models are combined to estimate long-term demographic consequences, fit naturally into this framework. TMLE  is characterized by a two-step procedure. First, it obtains an initial estimate of the data-generating distribution or the relevant part of this distribution needed to evaluate the target of inference. 
Second, TMLE updates this initial fit through a targeted step that optimally balances bias and variance specifically for the target of inference, rather than the entire distribution. This involves constructing a parametric submodel around the initial estimate and refining it to minimize bias for the target. 

TMLE is attractive in the context of IPMs because it provides a natural framework for uncertainty quantification of quantities explicitly obtained as consequences of models.  It is also attractive in being asymptotically efficient among unbiased models in a semiparametric framework \citep{van2018targeted}. It also retains robustness to poor initially estimates so long as nuisance parameters are estimated at a  $n^{-1/4}$ rate; in particular, it makes few other assumptions about plug-in estimators allowing us to leverage advanced machine learning tools and any other modeling strategy that best fits the data we have.


In this paper we introduce TMLE as a tool for ecological modeling, with the goal of developing robust and efficient estimators for several key parameters in IPMs: the long-term population growth $\lambda$ (the Perron-Frobenius dominant eigenvalue of the projection operator), the elasticity of $\lambda$ with respect to fecundity, and the average growth rate with regard to the year-specific covariate $\log \lambda_S$.  First we derive the efficient influence function for the specific target of inference within IPMs; we then design appropriate submodel paths tailored to the structures and settings of these ecological models. This process requires careful consideration of TMLE conditions and computational complexity. Potential challenges arise from the regularity conditions of TMLE for which we propose several solutions and discuss possible directions for future research. The main contribution of our work lies in not only creating a framework for applying TMLE in quantitative ecology but also highlighting its potential power in scientific fields beyond those where it is currently being used (causal inference in clinical trials and economics). We aim to raise general awareness and interest among researchers about the capabilities of TMLE and the potential for much broader applications.

The paper is organized as follows: Section \ref{sec:ipm} introduces the background of size-structured IPMs and discusses the challenges of model selection when estimating model parameters. In Section \ref{sec:tmle}, we explain the mathematical framework behind TMLE and demonstrate its effectiveness through asymptotic analysis. Section \ref{sec:tmle_ipm} presents the derivation of the influence function for IPMs and outlines the TMLE procedure for model updates. Section \ref{sec:sim} details our extensive experiments using both synthetic data and real data (Idaho and rotifer data). Finally, Section \ref{sec:dis} concludes the paper and suggests directions for future research.

\section{Size Structured IPM}
\label{sec:ipm}
\subsection{State Variable and Model Assumptions}\label{sec:ipm_assump}
In this section, we start by introducing the mathematical framework for IPMs. We focus on a simple size-structured IPM, in which the differences between individuals are completely summarized by a single attribute $z$ which is a continuous measure of body size such as total mass, volume, or longest leaf length. We use $n(z,t)$ to describe the size distribution at time $t$. We assume that $z$ takes values in a bounded interval $[L, U]$ representing the range of possible sizes. We assume that $n(z,t)$ is a smooth function of $z$ such that the number of individuals with size $z$ in the interval $[a,b]$ at time $t$ is $\int_a^bn(z,t)dz$. Note that $n(z,t)$ is not a probability distribution, in that its integral over the entire range of possible sizes represents the total population size rather than being exactly 1. The model operates in discrete time, going from times $t$ to $t+1$. The unit of time can be any duration such as one day, one year, etc. as appropriate for the species under study. Between times $t$ and $t+1$, individuals have the potential to either die or survive and (typically) change in size, they may also give birth to offspring that exhibit varying sizes.  

To describe these processes, we define $s(z)$ as the survival probability and $G(z',z)$ as the size transition kernel, meaning that $G(\bullet,z)$ is the probability density of size at time $t+1$ conditional on being size $z$ at time $t$. We use $F(z',z)$ to represent per-capita production of new recruits. So for small enough $h$, $G(z',z)s(z)h$ is the probability that the individual is alive at time $t+1$ and its size is in the interval $[z',z'+h]$. Similarly, $F(z',z)h$ is the number of new offspring in the interval $[z',z'+h]$ present at time $t+1$, per size-$z$ individual at time t. We use the kernel $K(z',z) = G(z',z)s(z)+F(z',z)$ to summarize the net result of survival and reproduction.
Then the population at time $t+1$ can be represented by 
\begin{equation}
n(z',t+1) = \int_L^U K(z',z)n(z,t)dz.
\label{eqn:defIPM}
\end{equation}
Note that the definition of kernel $K=Gs+F$ tacitly assumes a pre-reproductive census, meaning that reproduction occurs immediately after the population count at any time $t$, because it allows
 individuals alive at time $t$ to have descendents in the population at $t+1$ even if the individual does not survive to time $t+1$. The fecundity kernel $F$ then has to represent new
 offspring that survive until time $t+1$. Other census times (e.g., a census immediately after the breeding season) would imply different forms for the kernel.

Because kernel functions are generally not analytically integrable, an IPM is implemented using numerical integration methods based on evaluating the kernels and size-distributions at a finely spaced set of grid points. If numerical integrations are done using midpoint rule or higher-order 
``bin-to-bin'' quadrature methods \citep{2016:ellne.child.rees} the IPM is approximated by a matrix projection model (MPM) with a large number of narrow discrete size categories.  

When considering an MPM, we use $Z_t$ to denote the size class of an individual at time $t$, with $Z_t$ taking values in the finite set $\{1,2,\cdots,N\}$. In addition, we define the survival of an individual beyond time $t$ using a binary variable $S \in \{0,1\}$, where $0$ indicates that the individual dies before time $t+1$. Similarly, we denote the size class of an individual at time $t+1$ by the variable $Z^*_{t+1}$, which takes values in the set $\{0,1,2,\cdots,N\}$. Here, the value $0$ indicates death at time $t+1$,  while the rest correspond to the size class of survivors. We also incorporate fecundity into our model and define $Y_j$ to be the total number of offspring falling into the size class $j \in \{1,2,\cdots,N\}$ produced by this individual. Throughout this paper we will use the discretised version of an IPM in order to make our numerical calculations explicit.   Table \ref{table1} demonstrates an example of the data table for IPM. We see that for the individual with ID $3$, even though $Z^*_{t+1}$ is $0$ because of the death, we still observe the values of offspring size-classes $Y_j$.


 Mathematically, we introduce the following notations. We define the growth matrix $G$ by
\begin{equation}
G_{ji} := P(Z^*_{t+1}=j|Z_t=i,S=1) \qquad i,j=1,2,\cdots,N,
\end{equation}
which represents the size-class transition probability conditional on survival. 
The fecundity matrix $F$ denotes the expected number of offsprings in each size class produced by the individual, i.e.,
\begin{equation}
F_{ji} := E(Y_j|Z_t=i)  \qquad i,j=1,2,\cdots,N.   
\end{equation}
The survival matrix $M$ is defined by
\begin{align}
M := \text{diag}\Big\{P(S=1|Z_t=1),
\cdots,P(S=1|Z_t=N)\Big\},  
\end{align}
which is a diagonal matrix and each diagonal element corresponds to the conditional probability of survival given a particular size class.

For this paper, our interest is in the long term trend in population size predicted by the IPM. We start by considering two targets of inference for demonstration of TMLE. The first is the long term population growth defined by the dominant eigenalue of the combined matrices, i.e.,
\begin{equation}
\lambda = \text{max eigen}\Big( GM+F\Big).
\label{eqn:defLambda}
\end{equation}

The second is the total elasticity with respect to fecundity: 
\begin{align}
e = \Big [\frac{v^T F u}{\lambda \langle v, u \rangle} \Big ].
\end{align}
where $v$ and $u$ are the left and right dominant eigenvectors of $K = GM+F$ \citep{petersen2008matrix}. $\lambda$ is the maximum eigenvalue of $K$. This measures how fast the long term growth $\lambda$ would change if we proportionally perturb all the entries of the fecundity matrix by an infinitesimal amount (that is, $e$ is $\frac{1}{\lambda}\frac{d \lambda}{d \varepsilon}$ at $\varepsilon=0$ if $F$ is changed to $F + \varepsilon F, \varepsilon \ll 1$). This quantity is crucial for identifying which life stages or processes have the most significant impact on population dynamics. 

The above construction assumes survival, growth and fecundity all depend only on an individual's current size. We also consider the scenario where year-specific covariates exist (e.g. the Idaho plant population data in Section \ref{sec:sim}). Specifically, we define $\theta(t)$ as the random environment (or environmental covariate) in year $t$. The growth, survival, and fecundity matrices, conditional on the environmental covariate $\theta(t)$, are denoted as $G_{\theta(t)}, M_{\theta(t)}, F_{\theta(t)}$, respectively. The corresponding kernel matrix is defined as $K_{\theta(t)}$. In this case, the long term growth rate cannot be directly computed as the dominant eigenvalue of any single kernel or matrix in practice. Instead, we employ the small-fluctuation approximation, as described below, to address this challenge.
\begin{equation}
\log \lambda_S \approx E_{\theta(t)} \log \left( \frac{v^TK_{\theta(t)}u}{v^Tu} \right).
\label{eqn:SFA}
\end{equation}
where $v$ and $u$ represent the dominant left and right eigenvectors of the average kernel $\bar{K} = E_{\theta(t)} K_{\theta(t)}$, respectively. Further details can be found in Appendix \ref{appendix:lam}.
\begin{table}
 \caption{\label{table1}This is an example of the data table for IPM. The column "ID" identifies each unique individual. $Z_t$ denotes the size class of an individual at time $t$, while $S$ indicates whether the individual survived. $Z^*_{t+1}$ represents the size class at time $t+1$. Finally, $Y_j$ denotes the total number of offspring that fall into the size class $j$.} 
 \centering
\begin{tabular}{|c|c| c| c |c |c|c|c|} 
 \hline
 ID &   $Z_t$ & $S$& $Z^*_{t+1}$ & $Y_1$ & $Y_2$ & $\cdots$ & $Y_N$\\
 \hline
 1 &  6 & 1 & 21 & 0 & 1 &$\cdots$&0\\ 
 \hline
  2 &  11 & 1 & 34 & 2 & 1 &$\cdots$&0\\ 
 \hline
  3 &  23 & 0 & 0 & 1 & 1 &$\cdots$&0\\ 
 \hline
  4 &  5 & 1 & 32 & 0 & 0 &$\cdots$&0\\ 
 \hline
  5 &  33 & 1 & 52 & 1 & 0 &$\cdots$&0\\ 
 \hline
  $\cdots$  &  $\cdots$  & $\cdots$ & $\cdots$ & $\cdots$ & $\cdots$ &$\cdots$& $\cdots$\\ 
 \hline
\end{tabular}
\end{table}

\subsection{Model Selection}

To estimate the target of inference, the most straightforward approach is to separately estimate each component matrix ($G, F$ and $M$). Parametric models have traditionally been the standard approach for estimating these components, and this approach generally works well for $M$, where nonparametric regression can be easily employed if necessary. However, if we include the environment covariates, estimating $G$ and $F$ presents more challenges because one has to fit an entire size entire size distribution. Fitting size distributions parametrically can be challenging, especially when the underlying distributions are complex: multi-modal, or highly skewed. In such scenarios, parametric models may fail to capture the true structure of the data. Thus, there is a practical need for nonparametric approaches to estimate $G$ and $F$ more accurately (such as kernel regression or generalized additive models).
This inevitably involves model selection, for example choosing between linear and nonlinear parametric models or choosing the smoothing parameter in a nonparametric regression model. 
Traditional methods typically use cross-validation or a related criterion such as AIC to select the optimal model based on measures of prediction accuracy.  However, that approach does not focus on the particular target of inference. Model selection should be purpose-driven, targeting the specific question of interest. Moreover, as demonstrated through the asymptotic analysis in Section \ref{subsec:asym}, general parametric or nonparametric estimators may not exhibit desirable statistical properties yielding inconsistent uncertainty quantification for the target. In Section \ref{sec:demon}, we will use a simulation example to illustrate how hyperparameter tuning impacts the performance of estimates of the target of inference. Even with cross-validation for hyperparameter selection, there is no theoretical guarantee that the resulting estimator will exhibit desirable statistical properties, such as asymptotic normality. 

\section{Targeted Maximum Likelihood Estimation (TMLE)} \label{sec:tmle}
Briefly, TMLE is a statistical estimation method that targets the learning towards a target of inference and it is used to obtain influence statistics about the model parameters \citep{van2011targeted}. It provides robust and efficient estimation of target estimands by incorporating machine learning techniques and ensuring valid statistical inference. In this section, we will review the basic concepts of TMLE and discuss its implementation. For the convenience of the reader, we present a table of all the mathematical notations in Appendix \ref{appendix:math_notations}.

\subsection{Functional Mapping of Target of Inference} \label{subsec: funct}
Suppose that we observe a  independent and identically distributed sample 
$X_1, \cdots, X_n$ drawn from an unknown distribution $P$. We express our target of inference as being given by a functional $\Psi(P)$ of the data distribution $P$. Here, $\Psi$ is a known functional map from the distribution space to $\mathbb{R}$; in the context of IPMs the first of our targets is $\lambda$ written as a map from the distribution $P$ of observed $(Z_t, Z^*_{t+1}, S_t, Y_t)$. 
TMLE starts with a plug-in estimator for $\hat{P}$, yielding an initial estimate of the target $\Psi(\hat{P})$. And then develops a refinement of it for which we can provide inference. 

In some contexts, $\hat{P}$ can be as straightforward as the empirical distribution $P_n$ which places mass $1/n$ on each observed sample; classically, the mean function $\Psi_{\text{mean}}(P) = \int x dP(x)$ is naturally estimated by $\Psi_{\text{mean}}(P_n) = \int x dP_n(x) = \frac{1}{n} \sum X_i$. 
In more complex settings using the empirical distribution may be insufficient. In the context of an IPM, we can express the observed data distribution as 
\[
P(Z_t, Z^*_{t+1}, S_t, Y_t) = P(Z_t)G(Z^*_{t+1}|Z_t,S_t=1)M(S_t|Z_t)F(Y_t|Z_t)
\]
in which the calculation of $\lambda$ requires the values of $G$, $M$ and $F$ at values where we have no observed data. Here we plug in estimators $\hat{G}$, $\hat{M}$ and $\hat{F}$; within IPMs these are usually obtained from generalized linear models which may be subject to bias, or through nonparametric/machine learning methods for which uncertainty quantification is not available.  TMLE assumes a non-parametric specification of the initial estimate $\hat{P}$ but obtains uncertainty quantification from a  carefully designed parametric modification of these initial estimates, which we detail below. 


\subsection{TMLE Procedures and Motivation}\label{subsec:tmle_procedure}

We start by emphasizing some notation in order to clarify our exposition below. In particular, we will use four different representations of the underlying data distribution:
\begin{enumerate}[label=(\alph*), leftmargin=2em]
\item $P$: the theoretical target distribution where $\Psi(P)$ is our estimand.

\item $\hat{P}$: A plug-in estimate for $P$. This needs to approximate $P$ asymptotically; in practice it should be our best estimate of it through general parametric/ nonparametric methods.

\item $\tilde{P}$: A modification of $\hat{P}$ as described below so that $\Psi(\tilde{P})$ is asymptotically unbiased.

\item $P_n$: The empirical distribution of the data used to provide inference. This is assumed to be independent of $\hat{P}$, which we can guarantee by splitting the sample into data used to estimate $\hat{P}$ and those in $P_n$. This split needs to be proportional.
\end{enumerate}

In order to provide inference  for our target we must characterize the distribution of the errors that we make: 
\[
\sqrt{n}(\Psi(\hat{P}) - \Psi(P)) \sim D
\]
for some distribution $D$. This is challenging without specifying how we arrived at $\hat{P}$ and in particular $D$ may not have mean 0, meaning $\Psi(\hat{P})$ is asymptotically biased.  Instead, targeted learning proceeds in two steps:  
\begin{enumerate}[label=(\alph*), leftmargin=2em]
\item Obtaining a modification $\tilde{P}$ of $\hat{P}$ that is designed to remove the bias $\Psi(\hat{P})$ and is amenable to the next step. 

\item Finding an approximation to the error via an {\em influence function} $\psi(x,P)$. This is a function of the data $X$ that is defined with respect to a distribution $P$, and approximately represents the gradient of $\Psi$ with respect to perturbations of $P$.  We can then show that 
\begin{equation} \label{eq:asslin}
\sqrt{n} \Big [ (\Psi(\tilde{P}) -  \Psi(P))  - P_n \psi (X,P) \Big ]  \overset{p}{\to} 0 
\end{equation}
where $P_n \psi (X,P) := \frac{1}{n}\sum_{i=1}^n \psi(X_i,P)$ and notation $P_n$ is reused here to denote the empirical operator.  The derivation of $\psi$ from $\Psi$ is detailed below. The purpose of this approximation is that $P_n \psi(X,P)$ is a simple average and  we can use a central limit theorem
\begin{align}
\sqrt{n} P_n \psi(X) = \frac{1}{\sqrt{n}} \sum \psi(X_i,P) \rightarrow N(0,\sigma^2)    
\end{align}
to characterize the uncertainty in the bias-corrected point estimate $\tilde{P}$, thus providing valid inference.  
\end{enumerate}

As we demonstrate below, beyond the regularity of $\Psi$, the use of sample splitting means that we only require that $\hat{P}$ is a consistent estimator of $P$ to be able to provide valid inference without having to characterize the uncertainty in $\hat{P}$ itself. 

Both the goals above are obtained through the use of the efficient influence function which acts like the first term in a Taylor expansion.  Specifically, for distributions $P_1$ and $P_2$ we define the influence function to be the derivative of $\Psi$ starting from $P_1$ in the direction of $P_2$.  That is $\psi(x;P_1)$ satisfies
\begin{align} 
\left.\frac{d}{d \epsilon}  \Psi(P_1+\epsilon(P_2-P_1))\right|_{\epsilon=0} & =\int \psi(x;P_1)\{ dP_2(x) - dP_1(x) \} \label{eq:influence_def} \\
\int \psi(X,P_1) dP_1(X) & = 0 \label{eq:zerofinfluence}
\end{align}
where the second condition will be automatically satisfied for our construction of $\psi$ below. 
In the context of distributions over discrete-valued random variables $X$ taking values in in $\{1,\ldots,K\}$, setting $p_k = P(X = k)$, $k = 1,\ldots,K$, $\psi$ can be calculated explicitly by taking the derivative with respect to incrementing the probability of each discrete value in turn:
\begin{align} \label{eq:pointgateaux}
\left.\psi(k;\mathbf{p}) = \frac{d}{d \epsilon} \Psi( (1-\epsilon) \mathbf{p} + \epsilon \delta_k) \right|_{\epsilon=0}    
\end{align}
where $\delta_k$ is the indicator $x = k$. Note here that $(1-\epsilon) \mathbf{p} + \epsilon \delta_k$ always sums to 1 and hence $\Psi$ is always being regarded as a functional of a proper distribution and we refer to this sequence of models as a ``submodel-path''. From this, it also holds that 
\begin{align} \label{eq:inf_def_discrete}
\left.\frac{d}{d \epsilon}  \Psi((1-\epsilon) \mathbf{p}_1 + \epsilon \mathbf{p}_2)\right|_{\epsilon=0}   = \sum_k \psi(k; \mathbf{p}_1) (\mathbf{p}_2(k) - \mathbf{p}_1(k)),
\end{align}
the discrete version of (\ref{eq:influence_def}).   

When $X$ takes continuous values, we  follow \cite{hines2022demystifying} in deriving influence functions via a perturbation in the direction of a point-mass distribution at $x$: 
\begin{align}  \label{eq:continue_gateaux}
\psi(x;P) = \left .\frac{d}{d\epsilon}{ \Psi((1-\epsilon)P + \epsilon \delta_x) } \right|_{\epsilon=0}   
\end{align}
where $\delta_x$ indicates a point-mass at $x$. 
It is possible to find targets $\Psi$, with corresponding influence functions $\psi$ which satisfy \eqref{eq:influence_def} but for which \eqref{eq:continue_gateaux} is not well defined, for example if $\Psi$ is not defined for point-mass distributions.  However, \eqref{eq:continue_gateaux}
 will be sufficient for the targets of inference in this paper. In this case, the linearity of the derivative operator used in obtaining \eqref{eq:inf_def_discrete} can be extended to continuous spaces:
\begin{align}
 \int \frac{d}{d\epsilon} \Psi((1-\epsilon)P_1+\epsilon \delta_x) dP_2(x) =  \frac{d}{d \epsilon}  \Psi((1-\epsilon)P_1 + \epsilon P_2)
\end{align}
where $P_2$ integrates over the location of point-mass contamination $\delta_x$, yielding \eqref{eq:influence_def}. \eqref{eq:zerofinfluence} then holds from 
\begin{align}
\int \psi(x;P) dP(x) = 
\left. \frac{d}{d\epsilon} \Psi((1-\epsilon)P + \epsilon P) \right|_{\epsilon=0}= 0.
\end{align}
 At an intuitive level, $\psi(x;P)$ represents the direction within the space of distributions on which $\Psi$ is most sensitive, and therefore provides both a direction in which to correct bias and to quantify uncertainty.


Given a target $\Psi(P)$, its influence function $\psi(x;P)$ and an initial estimator $\hat{P}$, targeted learning first provides an update to $\hat{P}$.  For the sake of notational simplicity, we will assume that $\hat{P}$ admits a corresponding density $\hat{p}(x)$ and define a perturbed distribution:
\begin{align}
\hat{p}_{\epsilon}(x) = C(\epsilon)e^{\epsilon \psi(x;\hat{P})}\hat{p}(x)    
\end{align}
where $C(\epsilon)$ is a normalizing constant. Here  $\epsilon$ is treated as a parameter in $\hat{p}_{\epsilon}$ that is estimated by maximum likelihood using data that was held out when estimating $\hat{p}$. This yields a perturbation of $\hat{p}$ that has reduced bias:
\begin{gather}
\begin{split}
\epsilon^* & = \mbox{argmax} \sum \left[ \log  C(\epsilon) + \epsilon \psi(X_i;\hat{P}(X_i)) + \log \hat{p}(X_i) \right] \\
\tilde{p}(x) & = C(\epsilon^*)e^{\epsilon^* \psi(x;\hat{P})}\hat{p}(x)     
\end{split}   
\end{gather}

At this point we can obtain a new influence function $\psi(x;\tilde{P})$ and use it to further modify $\tilde{P}$. These updates are continued until maximum likelihood estimates yield no further adjustment: $\epsilon^* = 0$. 
Once our final $\tilde{P}$ is found, we obtain uncertainty quantification from its influence function via a confidence interval for $P_n \psi(X;\tilde{P})$; that is we rely on \eqref{eq:asslin} to to estimate a variance for $\Psi(\tilde{P})$ and use normal quantiles and tests for inference.   The entire procedure is summarized in Algorithm \ref{alg1}.

\begin{algorithm} 
\caption{TMLE}\label{alg1}
\begin{algorithmic}[1]
\State \textbf{Input:} Initial estimate $\hat{p}$ produced by a general machine learning method. 
\State Construct the sub-model path in the direction of the influence function:
\[
\hat{p}_{\epsilon}(x;\psi(x;\hat{P})) = C(\epsilon) \hat{p}(x)e^{\epsilon \psi(x;\hat{P})}
\]

\Repeat
    \State Perform maximum likelihood estimation to find $\hat{\epsilon}$:
    \[
    \hat{\epsilon}=\underset{\epsilon}{\arg \max } P_{n} \log \hat{p}_{\epsilon}
    \]
    \State Update the model:
    \[
    \hat{p}(x) \gets \hat{p}_{\hat{\epsilon}}(x;\psi(x;\hat{P}))
    \]
\Until{$\hat{\epsilon}$ converges to 0}
\State \textbf{Output:} $\Psi(\hat{P})$, with estimated standard deviation $\sqrt{ \frac{1}{n^2} \sum \psi(X_i,\hat{P})^2 }$.
\end{algorithmic}
\end{algorithm}

\subsection{A Justification of Targeted Learning} \label{subsec:asym}

Here we will provide an informal justification for the targeted learning framework, with a particular focus on motivating the procedure above.  To carry this out formally, we need two assumptions about the behavior of $\Psi$ and $\hat{P}$.  The first of these is asymptotic linearity which essentially states \eqref{eq:asslin} holds: 
\begin{assump} \label{assump1}
We define that the estimator $\Psi(\tilde{P})$ satisfies \emph{asymptotic linearity} if 
\begin{align} \label{Equ:assump1}
\sqrt{n} \Big [ (\Psi(\tilde{P}) -  \Psi(P))  - P_n \psi (X) \Big ]  \overset{p}{\to} 0 
\end{align} where  $\psi$ is some function such that $E_P \psi(X) = 0$ and has finite variance
\end{assump}
For our inference to be valid, we also require that this limiting distribution doesn't change depending on how $\hat{P}$ approaches $P$ for $P$ within a neighbourhood of $P$. This is stated formally as
\begin{assump} \label{assump2}
Pick $P_{\epsilon =  \frac{1}{\sqrt{n}}}:= P + \frac{1}{\sqrt{n}}(\tilde P-P)$.
Let $\tilde{\mathbb P}_n$ denote the empirical distribution of $n$ samples drawn from the perturbed distribution $P_{\epsilon = \frac{1}{\sqrt{n}}}$. An  estimator is regular if and only if, for all such such sequences we have 
    \begin{align}
    \sqrt{n} 
    \big[
    \Psi(\tilde{\mathbb P}_n) - 
    \Psi(\tilde P_{\epsilon =  \frac{1}{\sqrt{n}}})
    \big] 
    & \overset{P_{\epsilon =  \frac{1}{\sqrt{n}}}}{\rightsquigarrow}
    \mathcal D
    \end{align}
    Where $\mathcal D$ is some fixed distribution that doesn't depend on the choice of path $\tilde P_{\epsilon =  \frac{1}{\sqrt{n}}}$. 
\end{assump}
An target $\Psi(P)$ satisfying these assumptions is labelled regular and asymptotically linear (RAL).  We now need to construct the influence function $\psi$ and the bias corrected estimator $\tilde{P}$. 

We now motivate the targeted learning calculation.
The development of TMLE procedures is guided by a particular representation of the the error $\Psi(\hat{P}) - \Psi(P)$ which we derive here. As a starting point, we will center the influence function around $\hat{P}$ to write 
\begin{align}
    \Psi(P) - \Psi(\hat{P}) = \int \psi(x;\hat{P})(p(x)-\hat{p}(x))dx + R_2
\end{align}
and use property \eqref{eq:zerofinfluence} to replace $\int \psi(x;\hat{P})\hat{p}(x)dx = \int \psi(x;P) p(x) dx = 0$ resulting in an expression for the difference in the influence functions:
\begin{align}
\Psi(P) -  \Psi(\hat{P}) =  \int \Big(\psi(x;\hat{P}) -\psi(x;P) \Big)p(x) dx  + R_2 
\end{align}
Since $P$ is unknown, we will approximate it with $P_n$ and calculate the error
\begin{align}
\begin{split}
\Psi(\hat{P}) - \Psi(P)  & = - \int \Big(\psi(x;\hat{P}) -\psi(x;P) \Big)\Big(p(x)-p_n(x) +p_n(x)\Big) dx  - R_2 \\
&= - \int \Big(\psi(x;\hat{P}) -\psi(x;P) \Big)p_n(x) dx  \\ &  \hspace{1cm} + \int \left(\psi(x;\hat{P}) - \psi(x;P) \right)(p_n(x) - p(x))dx - R_2 \\
&= \frac{1}{n} \sum \psi(X_i;P) - \frac{1}{n} \sum \psi(X_i;\hat{P})  \\ & \hspace{1cm} + \int \left(\psi(x;\hat{P}) - \psi(x;P) \right)(p_n(x) - p(x))dx - R_2
 \end{split}
\end{align}

We note that the expansion above holds when replacing $\hat{P}$ with $\tilde{P}$. 
The first term above is the average of the "true" influence function (i.e., calculated at the true model), which has a mean of zero and follows a central limit theorem (because influence function is mean zero and has finite variance as defined in Assumption \ref{assump1}). In order to use this for inference, we need to ensure that the remaining three terms are negligible.  These are:

\begin{itemize}
\item \textbf{Plug-in Bias:} The term $ \frac{1}{n} \sum \psi(X_i;\hat{P})$ is known as plug-in bias. For a plug in estimate $\hat{P}$ this will not necessarily be negligible; but it is exactly the term that is set to zero in the perturbed model $\tilde{P}$ since when the update terminates
\[
0 = \left. \sum_i \frac{d}{d\epsilon} \log C(\epsilon)e^{\epsilon \psi(X_i;\tilde{P})}\tilde P(X_i) \right|_{\epsilon = 0} = \frac{nC'(0)}{C(0)} + \sum \psi(X_i;\tilde{P})
\]
and $C'(0) = \int \psi(x;\tilde{P}) \tilde{p}(x)dx = 0$. 

\item \textbf{Empirical Process:} The cross-product $ \int \left(\psi(x;\hat{P}) - \psi(x;P) \right)(p_n(x) - p(x))dx $ is known as the empirical process term, where we need to control the covariance between the two product terms.  In a sufficiently regular model -- e.g. under Donsker conditions \citep{shorack2009empirical} --  this term can be shown to be negligible even if we use the same data for $\hat{P}$ and $P_n$. However, if $P_n$ is  obtained independently from $\hat{P}$, say by sample splitting, this term has an expectation of zero, and as long as $\hat{P}$ converges to $P$, the term is $o_P(1/\sqrt{n})$, making it still negligible.
\item \textbf{Second-Order Remainder:} $R_2$ is the second-order remainder of the `Taylor expansion'. To control $R_2$, we require the regression methods used for the initial estimate $\hat{P}$ provide a sufficiently accurate estimate, such that $R_2$ can be controlled within an order of $o_P(1/\sqrt{n})$. In practice, we can use cross-validation for model selection and choose the model with the best predictive performance.
\end{itemize}

Combined these motivate the procedure in Algorithm \ref{alg1}; first obtain the best estimate $\hat{P}$ that you can on one sample, then use a second sample to both update it to remove the plug-in bias, and remove the covariance between $\hat{P}$ and our measurement of its variance.  The resulting framework allows us to discount the the effects of model selection, smoothing parameter choice etc in developing $\hat{P}$ (so long as we do not add so much variance that it becomes inconsistent) while still providing formal statistical uncertainty quantification for the target $\Psi(P)$. 

The {\em efficiency} of the influence function here is derived by considering any one-dimensional model $p_\epsilon(x) = C(\epsilon) e^{\epsilon h(x)}p(x)$ under a correct model $p$. For this model the MLE $\hat{\epsilon}$ has minimum asymptotic variance among all estimators, given by $1/[n E_P h^2(X)]$. Further, a delta-method argument gives the variance of $\Psi(P_{\hat{\epsilon}})$ as $E_P(\psi(X;P)h(X))^2/[n E_P h(X)^2]$ which is minimized when $h(x) = \psi(x;P)$ resulting in a minimum-variance estimate.  In the context of models with parametric components, model perturbations must stay within the space defined by the parametric family, meaning that $\psi(x;P)$ may no longer be the perturbation that results in smallest variance.

\section{TMLE for IPMs} \label{sec:tmle_ipm}
In this section, we will derive and apply the TMLE iteration in a novel application, estimation of the kernel function for integral projection models (IPM). We demonstrate application of TMLE to several different targets of inference: the long term population growth  $\lambda$ (dominant eigenvalue of the
projection kernel), the total elasticity of the kernel with respect to fecundity, and the average growth rate $\log \lambda_S$ under the environment covariates. It is important to note that TMLE can be broadly applied to various targets besides these for IPMs, and to other kinds of models in ecology. 

The first step in TMLE is to compute the influence function for the target of inference, which involves mathematical derivations using the chain rule according to the differential of the functional. We then use the influence function to construct the submodel path for the TMLE procedure. For convenience, we decompose the model space into two subspaces. Finally, we discuss the limitations and challenges of the proposed method and suggest potential solutions.

\subsection{Influence Function for IPMs}
 We denote the kernel function by $K = GM+F$ and use $u$ and $v$ to represent its left and right dominant eigenvectors. 
 Here we note that the full model $\hat{P}$ is a product of submodels $\hat{G}$, $\hat{F}$ and $\hat{S}$ and we can decompose the influence functions into a sum over each component and update each of these components separately. Note that our targets do not depend on the distribution of $Z_t$. 
 We present the influence functions for each target in the following two theorems.
 \begin{theorem}  \label{thm1}
Under the model setting in Section \ref{sec:ipm}, the influence function for the long term growth $\lambda$ can be computed by
\begin{align*}
\psi_{\text{Trend};Z^*_{t+1}|Z_t}(Z_{t},Z^*_{t+1},Y_1,\cdots,Y_N) = u^TW_{1}v \\
\psi_{\text{Trend};Y_1,\cdots,Y_N|Z_t}(Z_{t},Z^*_{t+1},Y_1,\cdots,Y_N) = u^TW_{2}v
\end{align*}
where
\begin{align*}
W_{1,ji} =\frac{I\{Z_t=i,Z^*_{t+1}=j\} - I\{Z_t=i\} P(Z^*_{t+1}=j|Z_t=i)}{P(Z_t=i)} 
\end{align*}
and
\begin{align*}
W_{2,ji} =\frac{I\{Z_t=i\}\Big(Y_j- E(Y_j|Z_t=i)\Big)}{P(Z_t=i)} 
\end{align*}
for $i,j = 1,2, \cdots, N$.
\end{theorem}

\begin{theorem} \label{thm2}
Under the model setting in Section \ref{sec:ipm}, the influence function for the elasticity of fecundity can be computed by
\begin{align*}
&\psi_{\text{Elasticity};Z^*_{t+1}|Z_t}(Z_{t},Z^*_{t+1},Y_1,\cdots,Y_N) \\ & = \frac{ \Big [ v^T \widetilde{W}_{1} u_{2} + v^T_{2} \widetilde{W}_{1} u\Big] \lambda -  c  \Big [ v^T \widetilde{W}_{1}u \langle v,u \rangle + \lambda v^T \widetilde{W}_{1}u_{1} + \lambda v^T_{1} \widetilde{W}_{1}u]}{\lambda^2 \langle v,u \rangle} \\
&\psi_{\text{Elasticity};Y_1,\cdots,Y_N|Z_t}(Z_{t},Z^*_{t+1},Y_1,\cdots,Y_N) \\ & = \frac{ \Big [ v^T \widetilde{W}_{2} (u_{2}+u)  + v^T_{2} \widetilde{W}_{2} u\Big] \lambda -  c  \Big [ v^T \widetilde{W}_{2}u \langle v,u \rangle + \lambda v^T \widetilde{W}_{2}u_{1}  + \lambda v^T_{1} \widetilde{W}_{2}u]}{\lambda^2 \langle v,u \rangle}
\end{align*}
where we use the same notations for $W_{1}$ and $W_{2}$ in Theorem \ref{thm1}, denote the Moore-Penrose inverse by $^+$ and 
\begin{align*}
v^T_{1} = v^T (\lambda I - K)^{+} \qquad  v^T_{2} = v^TF (\lambda I - K)^{+}  \\
u_{1} =  (\lambda I - K)^{+} u \qquad u_{2} =  (\lambda I - K)^{+} F u  \\
 c = \frac{v^T F u}{\langle v,u \rangle } \quad
\widetilde{W}_{1} = W_{1} - \frac{v^T W_{1}u}{\langle v,u\rangle} \quad \widetilde{W}_{2} = W_{2} - \frac{v^T W_{2}u}{\langle v,u\rangle}
\end{align*}
\end{theorem}

\begin{theorem} \label{thm3}
Under the model setting in Section \ref{sec:ipm} and when the environmental covariates $\theta(t)$ is considered, the influence function for the average growth rate under the environmental covariates can be computed by
\begin{align*}
&\psi_{\text{Elasticity};Z^*_{t+1}|Z_t,\theta(t)}(Z_{t},Z^*_{t+1},Y_1,\cdots,Y_N,\theta(t)) \\ & = \frac{ \Big [ v^T \widetilde{W}_{1} u_{2} + v^T W_{1,\theta(t)} u + v^T_{2} \widetilde{W}_{1} u\Big]  -  c_{\theta(t)}  \Big [  v^T \widetilde{W}_{1}u_{1} + v^T_{1} \widetilde{W}_{1}u]}{ c_{\theta(t)} \langle v,u \rangle} \\
&\psi_{\text{Elasticity};Y_1,\cdots,Y_N|Z_t,\theta(t)}(Z_{t},Z^*_{t+1},Y_1,\cdots,Y_N,\theta(t)) \\ & = \frac{ \Big [ v^T \widetilde{W}_{2} u_{2} + v^T W_{2,\theta(t)} u + v^T_{2} \widetilde{W}_{2} u\Big]  -  c_{\theta(t)}  \Big [  v^T \widetilde{W}_{2}u_{1} + v^T_{1} \widetilde{W}_{2}u]}{ c_{\theta(t)} \langle v,u \rangle} 
\end{align*}
where we use the same notations for $W_{1,\theta(t)}$ and $W_{2,\theta(t)}$ in Theorem \ref{thm1} but conditional on the environmental covariate $\theta(t)$, denote the Moore-Penrose inverse by $^+$ and 
\begin{align*}
v^T_{1} = v^T (\lambda I - \bar{K})^{+} \qquad   \qquad v^T_{2} = v^T K_{\theta(t)} (\lambda I - \bar{K})^{+}  \\
u_{1} =  (\lambda I - \bar{K})^{+} u \qquad u_{2} =  (\lambda I - \bar{K})^{+} K_{\theta(t)}  u  \qquad c_{\theta(t)} = \frac{v^T K_{\theta(t)} u}{\langle v,u \rangle } \\
\widetilde{W}_{1} = E_{\theta(t)}W_{1,\theta(t)} - \frac{v^T W_{1,\theta(t)}u}{\langle v,u\rangle} \quad \widetilde{W}_{2} = E_{\theta(t)}W_{2,\theta(t)}  - \frac{v^T W_{2,\theta(t)}u}{\langle v,u\rangle}
\end{align*}
\end{theorem}
We provide the derivations of the influence functions above in Appendix \ref{appendix:eif}. Notably, the influence function of $\lambda$ has a concise form. For each entry in the $W_{1}$ matrix, it corresponds to the influence function for the conditional distribution $P(Z^*_{t+1}|Z_t)$. The influence function for the model space $ Z^*_{t+1}|Z_t$ can be viewed as a linear combination of the influence functions for each $P(Z^*_{t+1} = j|Z_t = i)$. We note that the target $E(Z^*_{t+1}|Z_t)$ is not RAL for continuous valued $Z_t$, but that integrating it over a bounded function is. 

Similarly, the influence function for the space $Y_1, \cdots, Y_N|Z_t$ is strongly related to that for the conditional expectation $E(Y_j|Z_t)$. The influence function for elasticity is more complex due to the higher curvature of the functional, and the presence of $\lambda$ in the denominator increases instability in our experiments below. 

\subsection{TMLE procedure of IPM Model}
With the derived influence functions, we are ready to design the procedures for TMLE. We give the example when environmental covariate is not considered but the whole procedure can be easily generalized  to the case by conditional on $\theta(t)$. As noted above, the model space can be broken up into submodels and each submodel updated independently.  For the model space $\{Z^*_{t+1}|Z_t\}$, we construct the submodel path as discussed in Section \ref{subsec:tmle_procedure} by 
\begin{align}
\hat{p}_{\epsilon_1}(z^*_{t+1}|z_t) = C(\epsilon_1)e^{\epsilon_1\hat{\phi}_{z^*_{t+1}|z_t}}\hat{p}(z^*_{t+1}|z_t),    
\end{align}
where $\hat{p}$ is the transition probability from the matrix $GM$ and $C(\epsilon_1)$ is the normalizing constant. Then TMLE update the submodel path by maximizing the empirical log-likelihood function
\begin{align}
\hat{\epsilon}_1 = \arg\max_{\epsilon_1} P_n\log \hat{p}_{\epsilon_1}(Z^*_{t+1}|Z_t).    
\end{align}

For the model space $\{Y_1,\cdots,Y_d|Z_t\}$, since $Y_j$ takes values in an infinite set $\{0,1,2,\cdots\}$, we construct the sub model path in a different way. We use generalized linear models as working models \citep{van2011robust}. TMLE estimators are asymptotically unbiased and asymptotically normal, under mild regularity conditions. This means that even if the true data-generating distribution is not accurately described by a generalized linear model, the class of estimators we present will still be consistent and asymptotically normal. Moreover, if the generalized linear model used as the working model is correctly specified, the resulting estimator will be efficient and will achieve the semiparametric efficiency bound \citep{rosenblum2010simple}.

Specifically,  we consider a loss function that is similar to Poisson regression, i.e.,
\begin{align}
L(Q)(Z_t,Y_1,\cdots,Y_d) = \sum_j -Y_j\log Q_j(Z_t) + Q_j(Z_t),    
\end{align}
where $Q_j(Z_t) := E(Y_j|Z_t)$, which represents the $(i,j)$ entry of the $F$ matrix when $Z_t=i$. 
Then we can construct the submodel path
\begin{align}
\hat{Q}_{\epsilon_2,j} = \hat{Q}_{j} \text{exp}(\epsilon_2H_j).
\end{align}
where 
$$H_j(z_t) = \frac{\hat{u}_{z_t} \hat{v}_{j}}{\hat{P}(Z_t=z_t)}.$$
Then we solve the following optimization problem for TMLE update
\begin{align}
\hat{\epsilon}_2 = \arg\min _{\epsilon_2} P_n L(\hat{Q}_{\epsilon_2})(Z_t,Y_1,\cdots,Y_d).
\end{align}

As we discussed in Section \ref{subsec:asym}, there is a cross-product term that we need to control either by assuming the model is regular enough or using the sample splitting technique. In this paper, we use the idea of the cross-validated targeted minimum-loss based estimator (CV-TMLE) \citep{van2011cross}. It is a version of TMLE that uses sample splitting in order to make the TMLE maximally robust in its bias reduction
step. Mathematically, we define $B_n \in\{0,1\}^n$ to be a random vector indicating a split of $\{1, \ldots, n\}$ into a training and validation sample: $\mathcal{T}=\left\{i: B_n(i)=0\right\}$ and $\mathcal{V}=\left\{i: B_n(i)=1\right\}$. Let $P_{B_n, n}^\mathcal{T}, P_{B_n, n}^\mathcal{V}$ the empirical probability distributions of the training and validation samples, respectively. For a given cross-validation scheme $B_n \in\{0,1\}^n$, we now compute
\begin{align}
\hat{\epsilon}_1 &=  \arg \min _{\epsilon_1} E_{B_n} P_{B_n, n}^\mathcal{V} \log \hat{p}_{\epsilon_1}(P_{B_n, n}^{\mathcal{T}}),   \\
\hat{\epsilon}_2 &=  \arg \min _{\epsilon_2} E_{B_n} P_{B_n, n}^\mathcal{V} L\left(\hat{Q}_{\epsilon_2}\left(P_{B_n, n}^{\mathcal{T}}\right)\right).    
\end{align}
where the expectation $ E_{B_n}$ is taken with regard to the cross-validation scheme $B_n$. We summarize the procedure above in Algorithm \ref{alg2}. In practice, we split the sample into $V$ folds, with the index set denoted by $\mathcal{I} = {\mathcal{I}_1, \cdots, \mathcal{I}_V}$. For each $v \in {1, 2, \cdots, V}$, we treat the $v$-th fold $\mathcal{I}_v$ as the validation sample and the remaining folds $\mathcal{I}^{(-v)} = \mathcal{I} \setminus \mathcal{I}_v$ as the training sample. We use the superscript $(-v)$ to represent the model fitted on $\mathcal{I}^{(-v)}$, and $P^{(v)}_n$ represents the empirical average over the validation sample $\mathcal{I}_v$. The final TMLE estimate for the target of inference is the average of the estimates across all the validation folds. 

\begin{algorithm} \label{alg2}
\caption{Algorithm of TMLE for IPM}
\begin{algorithmic}[1]
\State \textbf{Input:} Initial estimate $\hat{G}^{(-v)}, \hat{M}^{(-v)}, \hat{F}^{(-v)}$ on each training sample for $v = 1,2,\cdots,V$. 
\State For each $v$, construct the sub-model path in the direction of the influence function:
\begin{align}
\hat{p}^{(-v)}_{\epsilon_1}(z_t,z^*_{t+1}) &= C(\epsilon_1)e^{\epsilon_1\hat{\phi}^{(-v)}_{z_t,z^*_{t+1}}}\hat{p}^{(-v)}(z^*_{t+1}|z_t),  \\
\hat{Q}^{(-v)}_{\epsilon_2,j} &= \hat{Q}^{(-v)}_{j} \text{exp}(\epsilon_2H^{(-v)}_j).
\end{align}

\Repeat
    \State Perform maximum likelihood estimation to find $\hat{\epsilon}_1$ and $\hat{\epsilon}_2$:
    \begin{align}
    \hat{\epsilon}_1 &= \arg\max_{\epsilon_1} \frac{1}{V} \sum_{v=1}^V P^{(v)}_n\log \hat{p}^{(-v)}_{\epsilon_1}(Z_t,Z^*_{t+1}), \\
    \hat{\epsilon}_2 &= \arg\min _{\epsilon_2} \frac{1}{V} \sum_{v=1}^V P^{(v)}_n L(\hat{Q}^{(-v)}_{\epsilon_2})(Z_t,Y_1,\cdots,Y_d).
    \end{align}
    \State Update the model:
    \begin{align}
    \hat{p}^{(-v)}(z_t,z^*_{t+1}) & \gets \hat{p}^{(-v)}_{\hat{\epsilon}_1}(z_t,z^*_{t+1};\psi(z_t,z^*_{t+1};\hat{P}^{(-v)})), \\
    \hat{Q}^{(-v)}_{j} &\gets \hat{Q}^{(-v)}_{j} \text{exp}(\hat{\epsilon}_2H^{(-v)}_j).
    \end{align}

\Until{$\hat{\epsilon}_1$ and $\hat{\epsilon}_2$ converges to 0}
\State \textbf{Output:} $\frac{1}{V} \sum_{v=1}^V \Psi(\hat{p}^{(-v)},\hat{Q}^{(-v)})$
\end{algorithmic}
\end{algorithm}
\subsection{Discussions}
Examining the efficient influence for both the growth and fecundity matrices reveals that they all contain the density of the size class (i.e., the term $P(Z_t=i)$) in the denominator. This is acceptable if the size classes are well-balanced and the density of each size class is significantly above zero. However, in practice, the real dataset may include some size classes with nearly zero density when a uniform grid is used to discretize the size values into different classes.  This is equivalent to the problems of positivity violation in the IPTW-estimator examined in  \citep{spreafico2024positivity}. 

In this paper, we propose creating the grid based on the quantiles of the size value distribution, rather than using a uniform grid for generating size classes. Mathematically, we define the set of split points by the true quantiles of $Z^c_t$, denoted as 
$\mathcal{I} = \{q_{1/N}, q_{2/N}, \cdots, q_{(N-1)/N}\}$, where $q_\alpha$ represents the $\alpha$ quantile of the data distribution. This method ensures that each size class is well-balanced.  The rationale behind this is that real data often includes a few very large size values. Using a uniform grid for size classes would result in many empty cells, which using quantiles automatically avoids. 

There are also alternative approaches that can be explored in future work.  For instance, we can penalize the loss function during the TMLE update. Mathematically, this can be achieved by adding an appropriate norm of the IPM matrices to the log likelihood criterion for the update. We could also consider a semi-parametric model and impose smoothness conditions on the data distribution. For example, we assume a linear relationship between $Z^*_{t+1}$ and $Z_t$. Here, even if $Z_t$ is missing in some size values, we can still extrapolate the relationship based on the smoothness assumption. However, this can result in model bias, and the caluclation of an influence function that is efficient within the model class can become quite intricate.


\section{Experiments} \label{sec:sim}
\subsection{Simulations} \label{sec:demon}
In this section, we begin with a straightforward case where no covariates are considered. This simplification will help us plot heatmaps of the kernel function and the efficient influence function, illustrating the differences under various hyperparameter choices and comparing initial estimates to TMLE-updated results. These visualizations will demonstrate how TMLE updates modify the original model. We will start by introducing the simulation setup, followed by the presentation and discussion of the results.

\subsubsection{Simulation Setup} \label{sec:sim_setup}
\begin{itemize}
    \item \textbf{Model Definitions} We first define $Z^c_t$ and $Z^c_{t+1}$ as the continuous size values at time $t$ and $t+1$ before they are discretized into size classes.  Here, $Z^c_t=0$ represents the seedling stage of the species. The values $Z^c_t$ and $Z^c_{t+1}$ are then categorized into size classes  $Z_t$ and $Z^*_{t+1}$. Following the same notations used in Section \ref{sec:ipm_assump}, we denote $S$ as the survival of an individual beyond time $t$, where 0 indicates the death. $Y_j$ represents the total number of offspring falling into the size class $j$ produced by this individual. 
    \item \textbf{Data generating process}: We generate the size values at time $t$ using the distributions: $Z^c_{t} \sim  (1-B) \text{Beta}(2,2)$, where $B \sim \text{Bern}(0.35)$. This indicates that 35\% of the samples are seedlings.  The size value at time $t+1$ follows a linear relationship: $Z^c_{t+1} \sim 0.8Z^c_{t}+ 0.2\text{Beta}(8,8)$.  We construct the grids with $100$ split points corresponding to the percentiles of the distribution of the size value $Z^c_t$. This enables the transformation of $Z^c_t$ and $Z^c_{t+1}$ into size classes $Z_t$ and $Z^*_{t+1}$. We consider model the survival by $S=\text{plogis}(0.1+7Z^c_{t})$. 
   For fecundity, we first use the model $Y \sim \text{Pois}\Big(e^{-3 + Z^c_t}\Big)$  to produce the offspring for the current individual. Then we randomly assign these offspring to each size class in proportion to  $p^{(f)} = (0.9,0.01,0.01,\cdots,0.01,0,\cdots,0)$, where $p^{(f)}_i$ represents the proportion of offspring that fall into the size class $i$; that is while most offspring are classed as seedlings, 10\% are spread uniformly across the first 10\% of the size range. 
    \item \textbf{Model fitting}: The procedure involves the following steps:
    \renewcommand{\labelenumi}{\theenumi)}
    \begin{enumerate}
        \item To construct the growth matrix, we first fit a linear regression model by regressing $Z^c_{t+1}$ on $Z^c_t$. The residuals from this linear model are denoted as $R_t$. The next step involves estimating the distribution of the residuals $R_t$. We apply kernel density estimation using various bandwidths ($0.01,0.02,0.03,0.05,0.01$)  as well as using cross-validation to select the optimal bandwidth. The growth matrix will be developed using both the linear model and the estimated kernel function.
        \item For survival, we use a logistic regression model to estimate the probability of survival. Specifically, we regress $S$ on $Z^c_t$, and the fitted model is then utilized to construct the survival matrix.
        \item For fecundity, we use two distinct Poisson regression models to estimate the expected number of offspring that become seedlings or non-seedlings. Specifically, the first Poisson model regresses the total number of seedlings $Y_1$ on $Z^c_t$. The second model regresses the non-seedling offspring $Y_j$ on both $Z^c_t$ and $Z_{r}$, where  $Z_r$ represents the size value corresponding to the size class $j$ into which the offspring is categorized. These Poisson models are then used to construct the fecundity matrix.
    \end{enumerate}
    \item \textbf{Parameters configurations}: We set the sample size to $1000$ and replicate the simulations 
    $200$ times using different random seeds. The sample is divided into 5 folds to implement CV-TMLE. We consider the two cases when the target of inference is long term growth $\lambda$ or elasticity. 
\end{itemize}

\subsubsection{Simulation Results}
\begin{figure}[t]
    \centering
    \includegraphics[width=\textwidth]{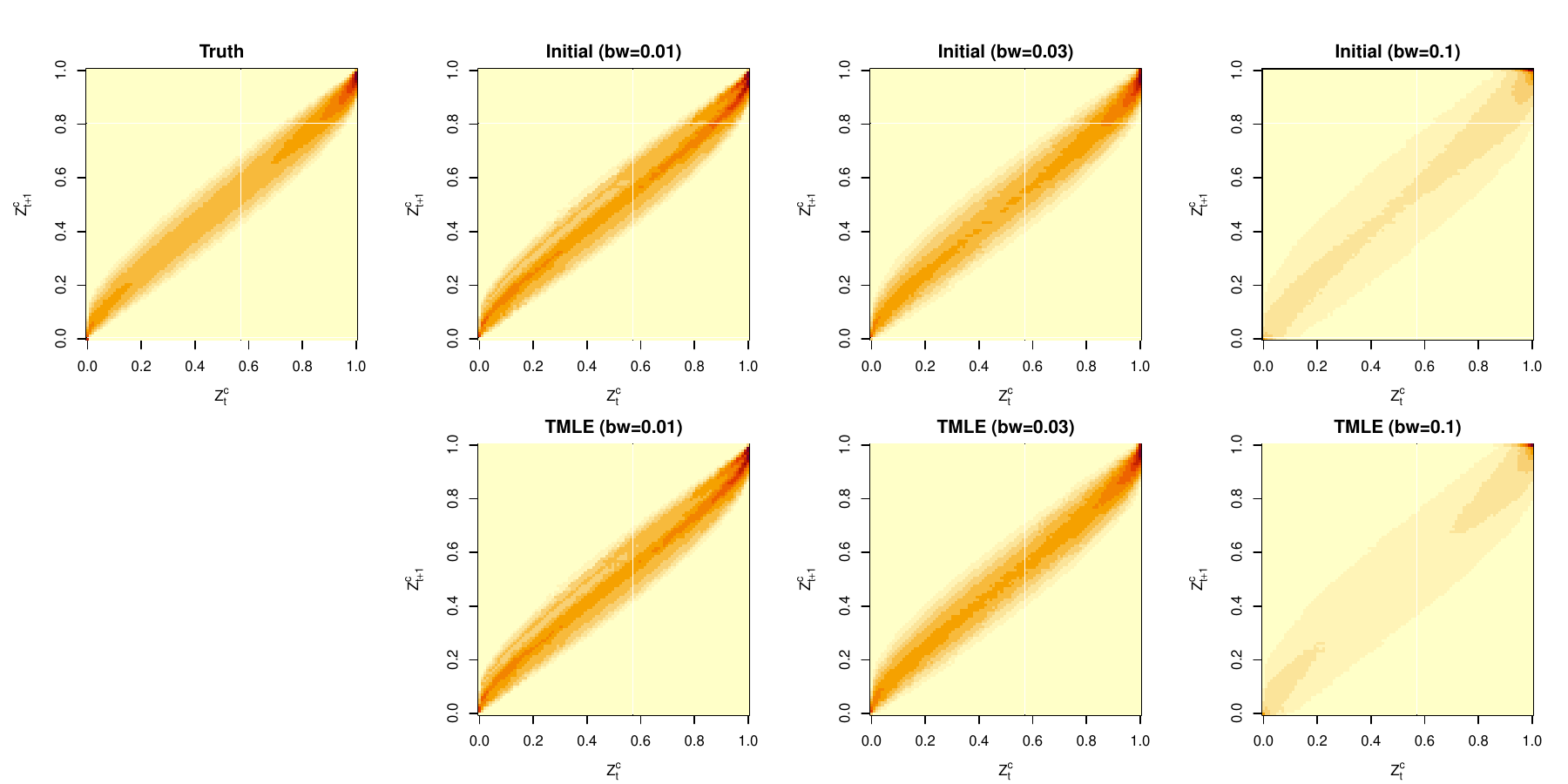}
    \caption{Heatmap for the matrix multiplication $MG$ when the target of inference is $\lambda$. We use the true matrix as the baseline for comparison. The term 'Initial' refers to the initially estimated matrix without any bias correction by TMLE. 'TMLE' indicates the results after applying the TMLE update. We present results for different bandwidths (bw) of $0.01, 0.03$ and $0.1$. }
    \label{fig:sg_lam}
\end{figure}

\begin{figure}[t]
    \centering
\includegraphics[width=\textwidth]{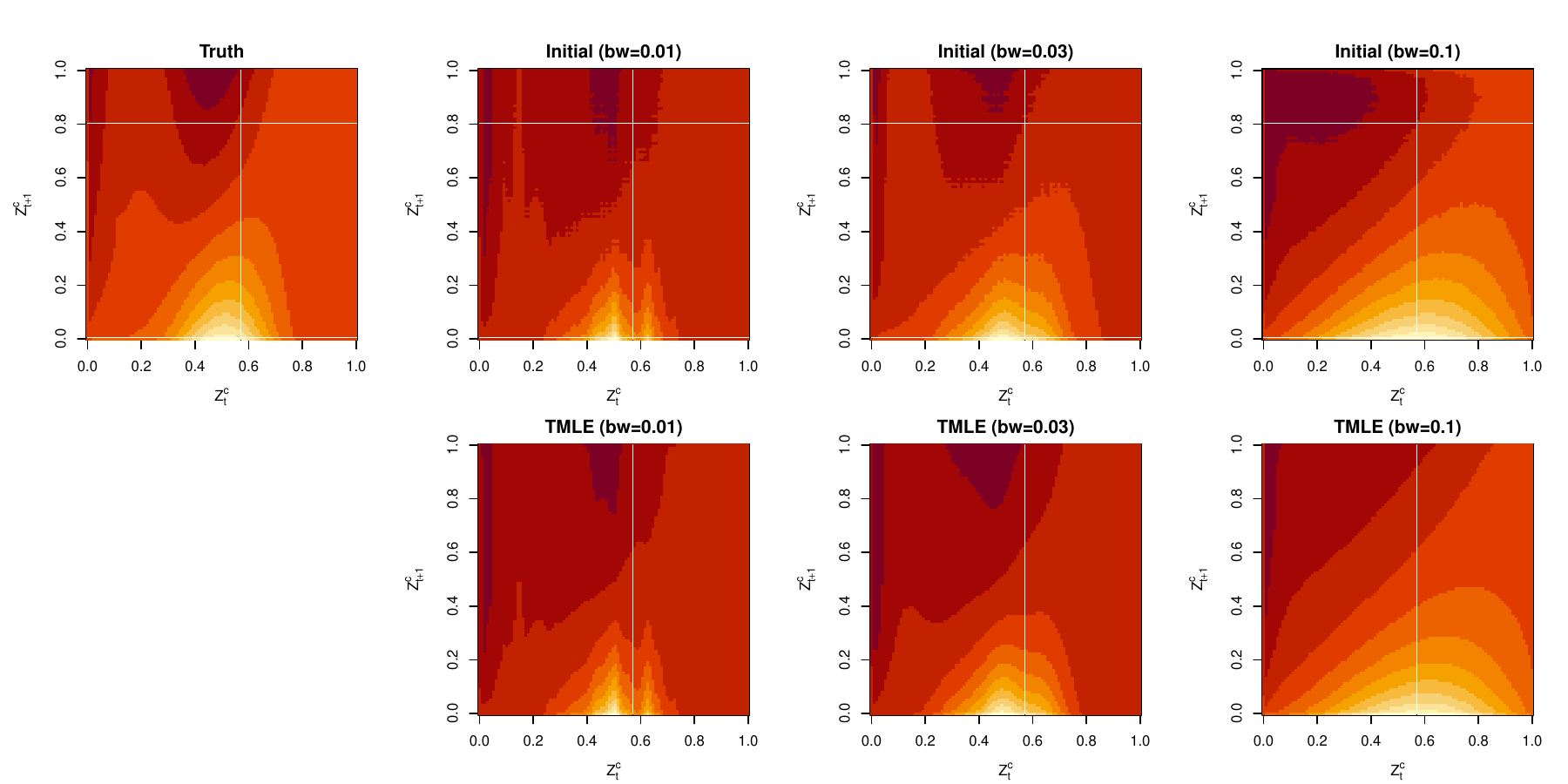}
    \caption{Heatmap for the influence function of $MG$ when the target of inference is $\lambda$. We use the true matrix as the baseline for comparison. The term 'Initial' refers to the initially estimated matrix without any bias correction by TMLE. 'TMLE' indicates the results after applying the TMLE update. We present results for different bandwidths (bw) of $0.01, 0.03$ and $0.1$. }
    \label{fig:eig_lam}
\end{figure}

When the target of inference is 
$\lambda$,  we first generate the heatmap plot for the matrix multiplication  $GM$ under different bandwidths, as shown in Figure \ref{fig:sg_lam}. We first observe that when the bandwidth increases, the model tends to over smooth the growth kernel, leading the kernel to have dramatic difference from the truth. We observe that as the bandwidth increases, the model tends to oversmooth the growth kernel, causing significant deviations from the true kernel. Although there is no obvious difference in the growth kernel before and after the TMLE updates, a small pattern appears where TMLE tends to shift the distribution of the kernel to the lower left and upper right corners, especially when bw = 0.1. In Figure \ref{fig:eig_lam} we plot the heatmaps of the estimated efficient influence functions; as the bandwidth increases, the shape of the influence function deviates from the truth. There are small differences in the patterns of the influence function before and after the TMLE updates. With the TMLE update, the influence function tends to become smoother and more closely resembles the true shape. 

\begin{figure}[t]
    \centering
\includegraphics[width=\textwidth]{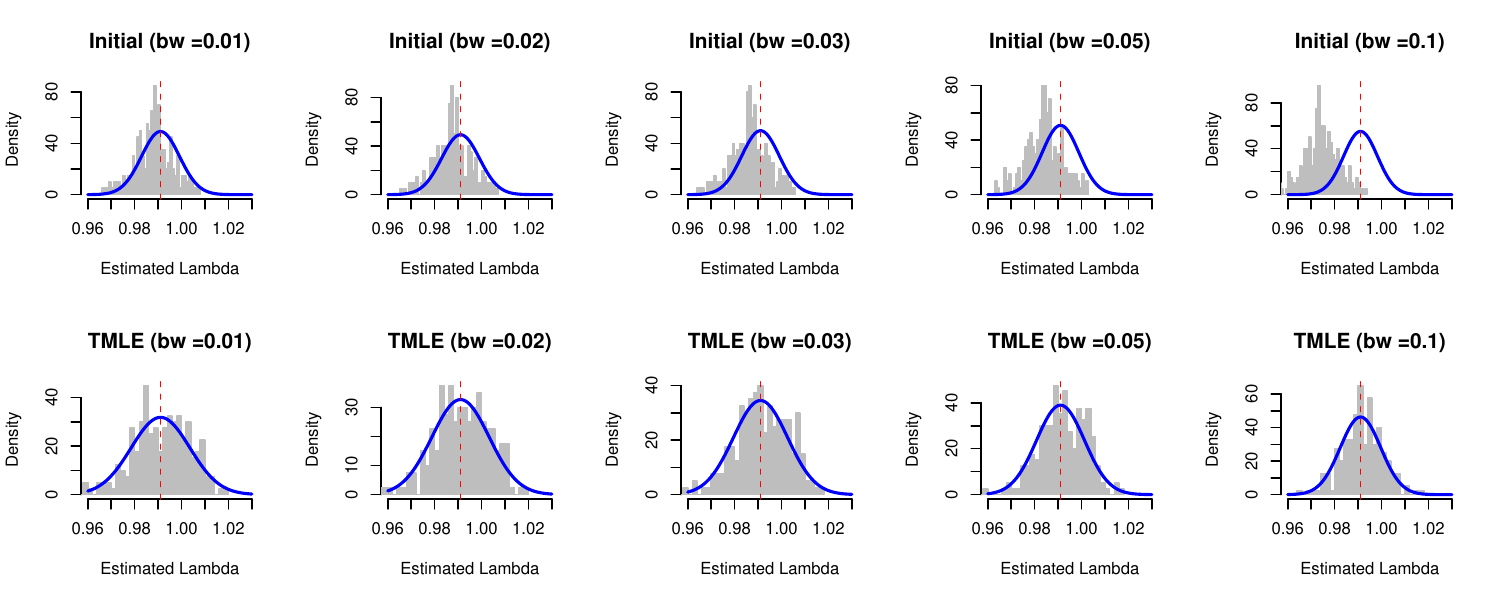}
    \caption{Histogram for the estimated $\lambda$ across 200 repetitions. The term 'Initial' refers to the initially estimated matrix without any bias correction by TMLE. 'TMLE' indicates the results after applying the TMLE update. We present results for different bandwidths (bw) of $0.01, 0.02, 0.03, 0.05$ and $0.1$.  The blue curve represents the density of a normal distribution with the true mean and the standard deviation of the estimators.}
    \label{fig:den_lam}
\end{figure}

\begin{figure}[t]
    \centering
\includegraphics[width=\textwidth]{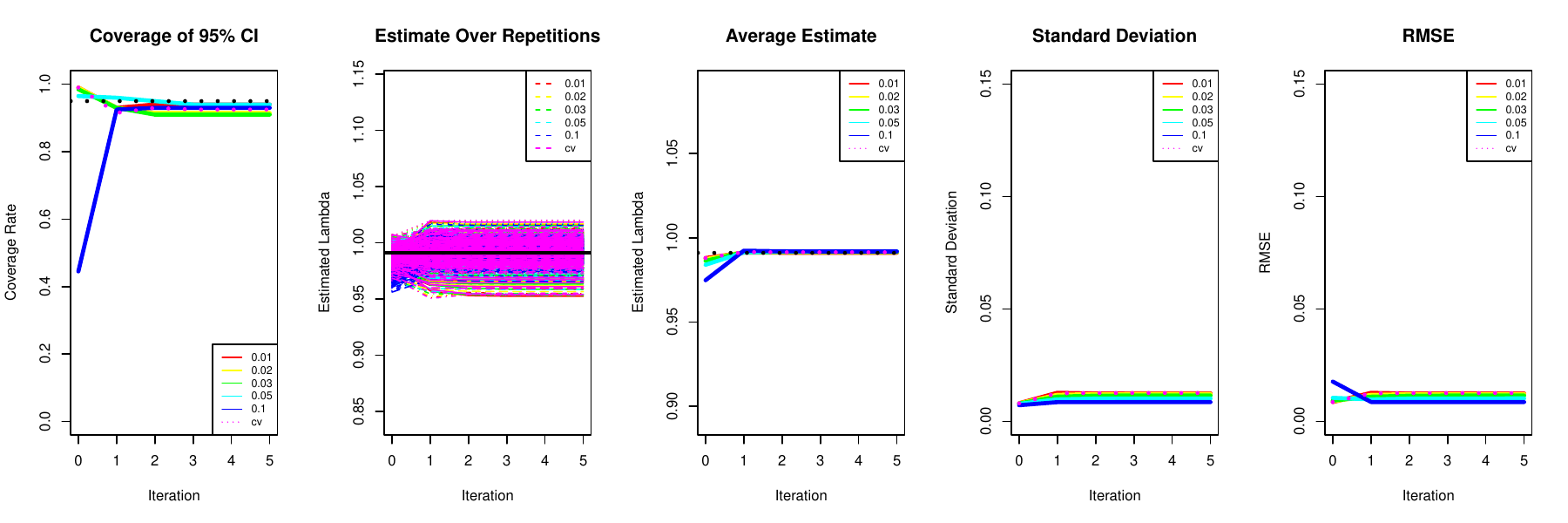}
    \caption{Results of the TMLE iterations for the target $\lambda$. We consider 5 iterations, with 0 representing the initial estimate. Different bandwidths (bw)  of $0.01, 0.02, 0.03, 0.05$ and $0.1$ are evaluated, including the optimal bandwidth selected by cross-validation (cv). We present results for the coverage of the 95\% confidence interval, estimates across repetitions, the average estimate, the standard deviation, and the root mean square error (RMSE). The simulation is repeated 200 times. }
    \label{fig:lam}
\end{figure}

Figure \ref{fig:den_lam} presents a histogram for the estimated $\lambda$ across 200 simulated data sets. The blue curve represents the density of a normal distribution with the true mean and the standard deviation of the estimators. We observe that for the initial estimate, as the bandwidth increases, the distribution of the estimators deviates further from the true value, and for all bandwidth values, the histogram is left-skewed. However, after the TMLE update, the bias in the distribution is removed, and the histogram adopts a more symmetric shape, closer to a normal distribution. In Figure \ref{fig:lam}, we present the results of the TMLE iterations for the target  $\lambda$. We consider 5 iterations, with 0 representing the initial estimate. Different colors are used to represent various bandwidth values. From left to right, the first figure shows the coverage of the truth for different estimates at the $95\%$ significance level. The second figure plots the estimates across repetitions, and the third figure shows their average, illustrating the bias of the estimates. The fourth figure displays the standard deviation of each method, and the fifth figure presents the root mean square error (RMSE). After the TMLE updates, we observe a robust and efficient estimate. The initial estimate suffers from either under-coverage or over-coverage depending on the bandwidth, but the TMLE iterations achieve the ideal 95\% coverage. Additionally, the bias in the estimate is eliminated, bringing the average estimate back to the true value. Similarly, we also represent the results when the target of inference is elasticity from Figures \ref{fig:sg_ela} to \ref{fig:ela} in the Appendix.

\subsection{Idaho Data}\label{subsec:idaho}
\subsubsection{Dataset Description}
In this section, we consider the idaho data \citep{adler2018weak}. The study conducted at the U.S. Sheep Experiment Station, located near Dubois, Idaho, focuses on the interactions between dominant plant species in a sagebrush steppe ecosystem. The primary aim was to investigate the competitive dynamics and stability of species coexistence through both observational data and experimental manipulation. The vegetation is predominantly composed of four shrub species: Artemisia tripartita (ARTR) and the C3 perennial bunchgrasses Hesperostipa comata (HECO), Pseudoroegneria spicata (PSSP), and Poa secunda (POSE). Specifically, the dataset includes historical plant data collected between 1926 and 1957. During these periods, 26 1-m² quadrats were monitored annually, tracking individual plants' survival, growth, and recruitment using a pantograph mapping method.  Eighteen quadrats were placed within four ungrazed exclosures, and eight quadrats were placed in two paddocks grazed at medium intensity from spring through fall. All quadrats were situated on similar topography and soils \citep{adler2018weak}. The data is publicly available on on the \href{https://datadryad.org/stash/dataset/doi:10.5061/dryad.96dn293}{Dryad Digital Repository}  \citep{Adler_2019}.

\subsubsection{Data Processing}
The dataset includes annual survival and growth data for each individual plant. 
We treat the year as a categorical factor and include it in the baseline covariates. We apply a log transformation to the original size values of each individual and then rescale them to a [0,1] range. These values are then discretized into size classes by creating 100 bins based on their percentiles. Recruitment data is available at the quadrat level. To allocate recruitment to individual plants, we first identify the individuals within each quadrat. Recruitment at time $t+1$is then assigned to each individual in proportion to the square root of their size at time $t$ corresponding to  the pre-breeding design as mentioned in Section \ref{sec:ipm_assump}. This method allows us to format the dataset according to the model structures outlined in Section \ref{sec:ipm_assump}. Additionally, extra datasets provide climate information, such as \texttt{ppt1}, \texttt{ppt2} (total precipitation over the past two years), and \texttt{TmeanSpr1}, \texttt{TmeanSpr2} (mean spring temperature over the past two years) and \texttt{totParea} (total area currently covered by the species). These are included as environmental covariates. Because of these year-specific covariates, we consider the average growth rate as the target of inference $\log \lambda_S$ in this experiment. Further details on the data processing procedure are provided in Appendix  \ref{appendix:idaho_proc}.

\subsubsection{Synthetic Data Analysis} \label{sec:syn_analysis_idaho}
We start with a synthetic data analysis using HECO data as an example, which allows us to easily assess the performance of TMLE.  To generate the synthetic data, we consider a mixture distribution for $Z^c_t$.  First, we determine the proportion of $Z^c_t$ that takes the value zero (i.e., seedlings), and then a beta distribution is used to generate the non-seedlings. The covariates are generated using the bootstrap method. For growth, a linear model is employed to generate $Z^c_{t+1}$ based on $Z^c_{t}$ and other covariates, with the residual modeled using a scaled beta distribution.  Residuals are modeled with a scaled beta distribution. For survival, two separate logistic regression models are applied to seedlings and non-seedlings. Fecundity data at the quadrat level is generated using a Poisson model, and each offspring is randomly assigned to a size class according to the size class distribution within the total recruit population. All parameters in these models are fitted based on the real data. Detailed procedures for synthetic data generation are provided in the Appendix \ref{appendix:syn_idaho}. To construct the initial estimate, we begin with the growth matrix by fitting a linear regression model based on the current and subsequent size values for the surviving individuals. We then fit a beta distribution to the residuals. For survival, we employ logistic regression to fit two separate models: one for seedlings and another for non-seedlings. Fecundity is estimated using two distinct Poisson regression models, predicting the expected number of offspring that become seedlings or non-seedlings. As we are considering the pre-reproductive census described in Section \ref{sec:ipm_assump}, we should also account for the fecundity of individuals that die in time $t+1$. We use the same settings for the rest parameters configurations in Section \ref{sec:sim_setup}. For more details on the modeling process, please refer to the Appendix \ref{appendix:model_idaho}.

\begin{figure}[t]
    \centering
\includegraphics[width=\textwidth]{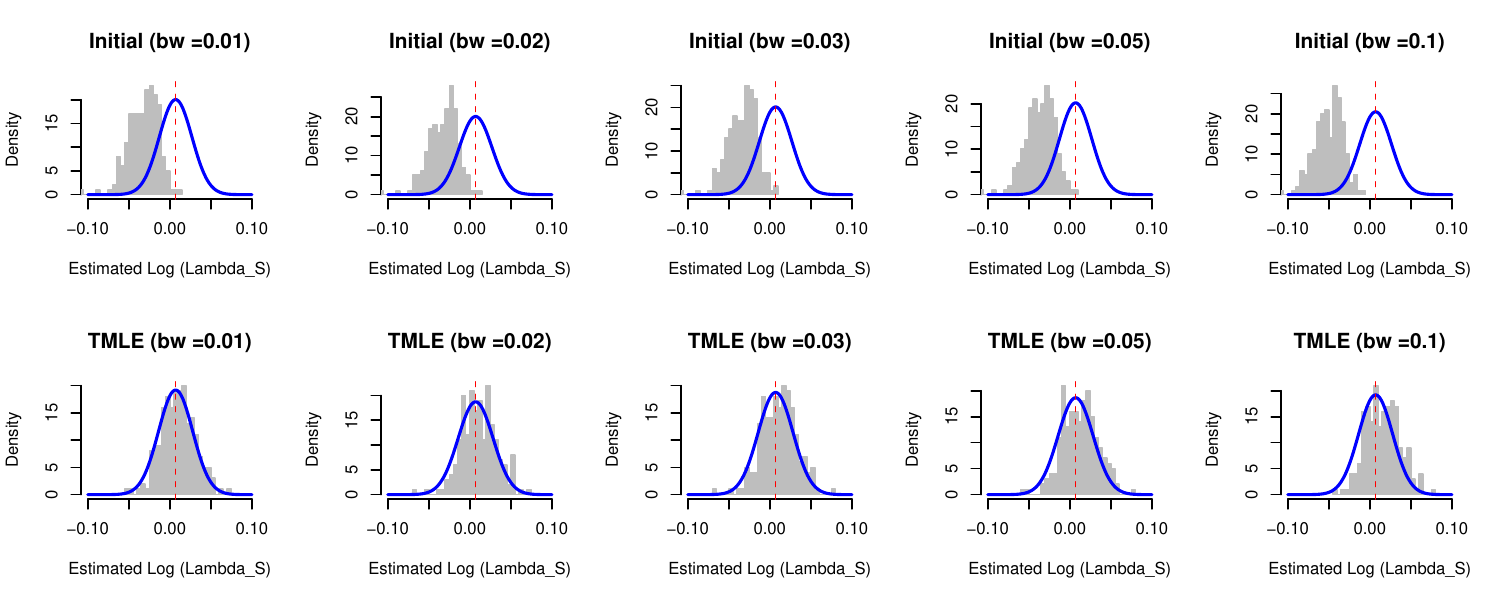}
    \caption{Histogram for the estimated $\log \lambda_S$ across 200 repetitions of HECO synthetic data. The term 'Initial' refers to the initially estimated matrix without any bias correction by TMLE. 'TMLE' indicates the results after applying the TMLE update. We present results for different bandwidths (bw) of $0.01, 0.02, 0.03, 0.05$ and $0.1$.  The blue curve represents the density of a normal distribution with the true mean and the standard deviation of the estimators.}
    \label{fig:den_lam_heco}
\end{figure}

\begin{figure}[t]
    \centering
\includegraphics[width=\textwidth]{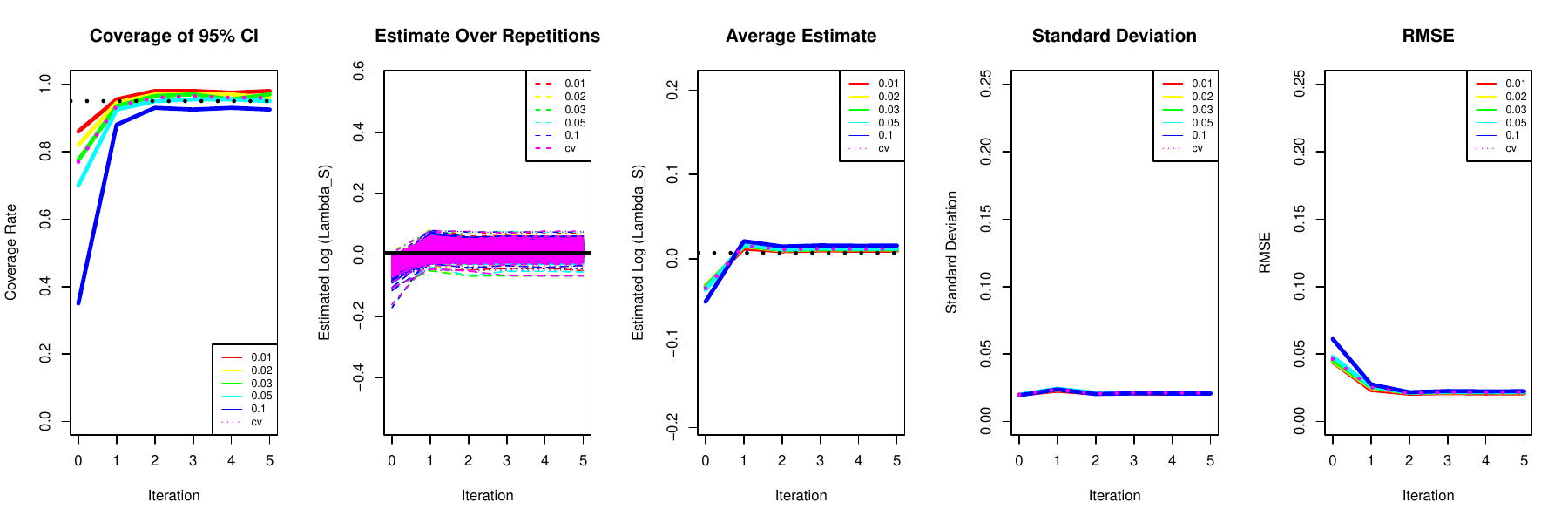}
    \caption{Results of the TMLE iterations for the target of inference $\log \lambda_S$ when the HECO data is considered. We consider 5 iterations, with 0 representing the initial estimate. Different bandwidths (bw)  of $0.01, 0.02, 0.03, 0.05$ and $0.1$ are evaluated, including the optimal bandwidth selected by cross-validation (cv). We present results for the coverage of the 95\% confidence interval, estimates across repetitions, the average estimate, the standard deviation, and the root mean square error (RMSE). The simulation is repeated 200 times. }
    \label{fig:hec_lam}
\end{figure}

We first present the histograms of the estimated $\log \lambda_S$ for the HECO synthetic data in Figures \ref{fig:den_lam_heco}. Similar to the findings in Section \ref{sec:demon}, the TMLE update shifts the distribution towards the true values, resulting in a more symmetric shape that is closer to a normal distribution. The results of TMLE iterations are shown in Figures \ref{fig:hec_lam}. We observe that TMLE updates yield robust estimates across different choices of the hyperparameter bandwidth. The method achieves good coverage and significantly reduces the bias of the estimates. Using the cross-validated bandwidth as the initial estimate, we achieve optimal coverage, and the final estimate is nearly unbiased. 

\subsubsection{Real Data Analysis}
\begin{figure}[t]
    \centering
\includegraphics[width=\textwidth]{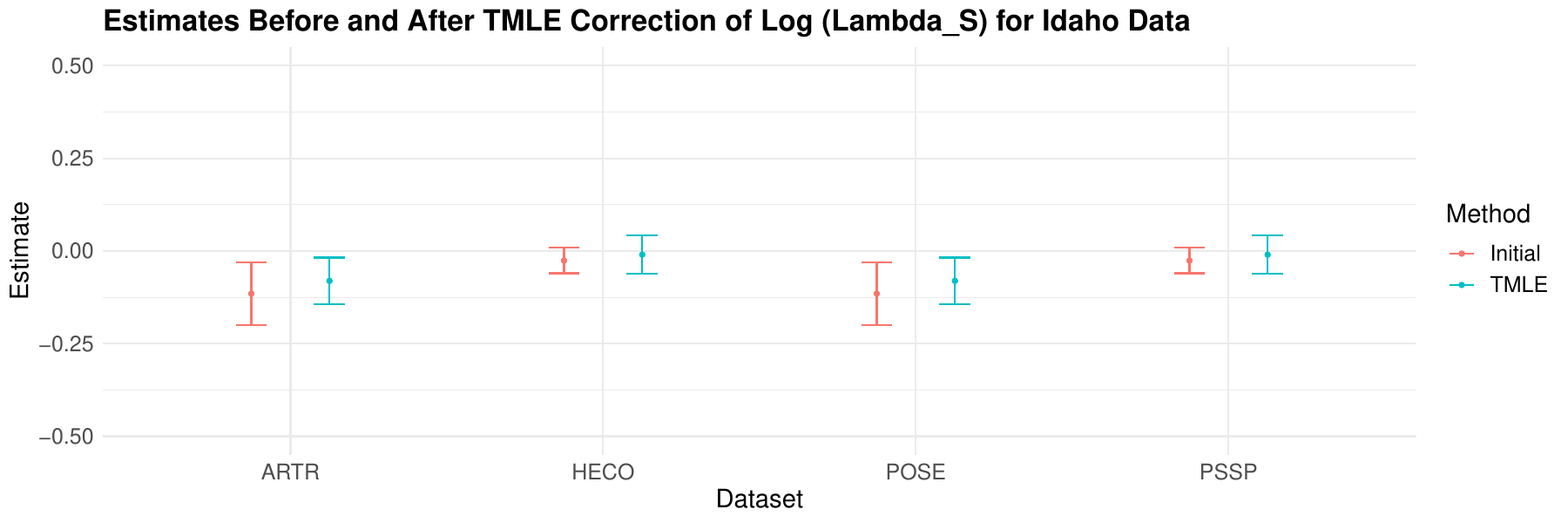}
    \caption{Real Data Analysis for Idaho Data. We present the estimate for $\log \lambda_S$   before and after the TMLE correction, along with the estimated confidence interval. }
    \label{fig:ci_real_lam}
\end{figure}

We conduct experiments on the real data for these four species. The cross-fitting technique is employed by splitting the dataset into five folds. The initial estimate is constructed using the same procedure described in Section \ref{sec:syn_analysis_idaho}. For each dataset, we present the estimate  before and after the TMLE correction, along with the estimated confidence interval in Figures \ref{fig:ci_real_lam}. We observe that the initial estimate tends to underestimate $\log \lambda_S$  according to the TMLE correction. 

\begin{figure}[t]
    \centering
\includegraphics[width=\textwidth]{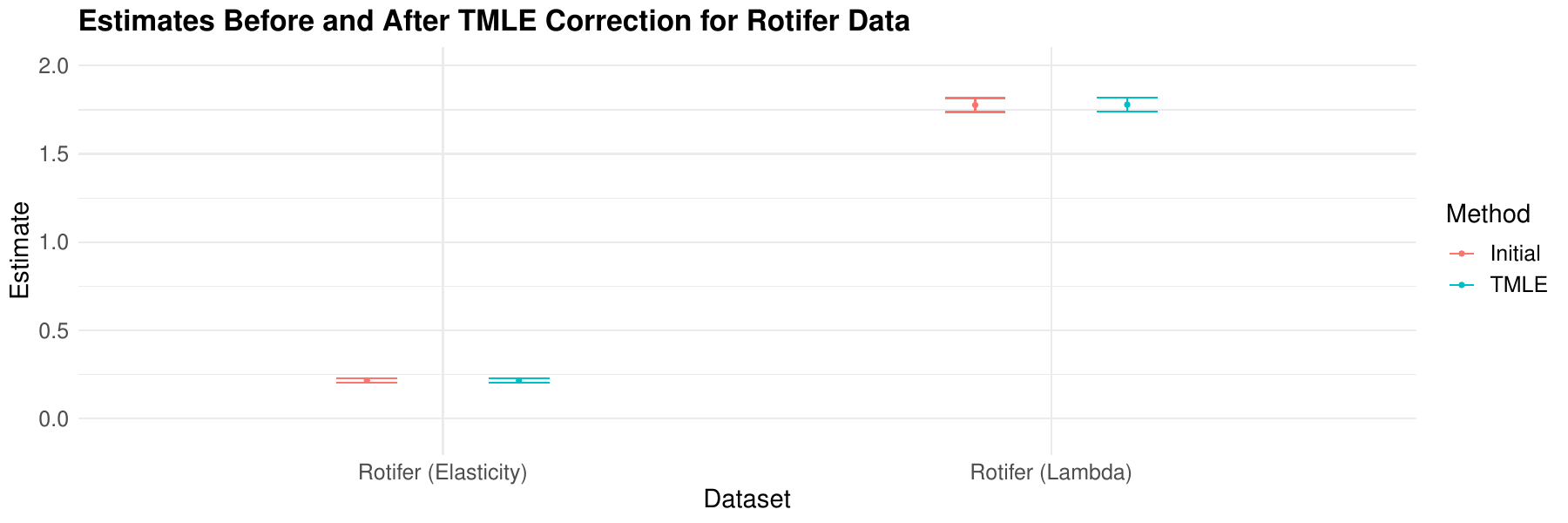}
    \caption{Real Data Analysis Rotifer Data. We present the estimate for $\lambda$ and elasticity   before and after the TMLE correction, along with the estimated confidence interval. }
    \label{fig:ci_real_rotifer}
\end{figure}

\subsection{Rotifer Data}
\subsubsection{Dataset Description}
We consider the Rotifer data in \citep{bock2019maternal}. The study focuses on maternal effect senescence in the aquatic invertebrate Brachionus manjavacas, investigating how maternal age impacts offspring survival and fertility. Each female produces 25 to 30 large daughters over about 10 days, with direct development and no post-hatching parental care, ensuring maternal age effects are not influenced by maternal care. Individual rotifers were housed and monitored for frequent, high-replication measurements of lifespan and fecundity.

\subsubsection{Data Processing}
We disregard the time dependence and treat each entry in the data table as an independent and identically distributed (i.i.d.) sample unit. We only include data from the first 16 days, as most rotifers die after this period, resulting in 16 age classes corresponding to each day. To simplify, we construct four maternal age groups for the offspring corresponding to coming from mothers aged 1, 2, 3, or 4, for maternal age group 1-4 d,  5 and 6 days for maternal age group 2, those aged 7 and 8 daysfor group 3, and those aged 9-16 days for group 4. We follow a similar procedure with \cite{hernandez2020demographic} to formalize the matrix system. We provide more details in Appendix \ref{appendix_rotifer_process}. 

\subsubsection{Synthetic Data Analysis}
\begin{figure}[t]
    \centering
\includegraphics[width=\textwidth]{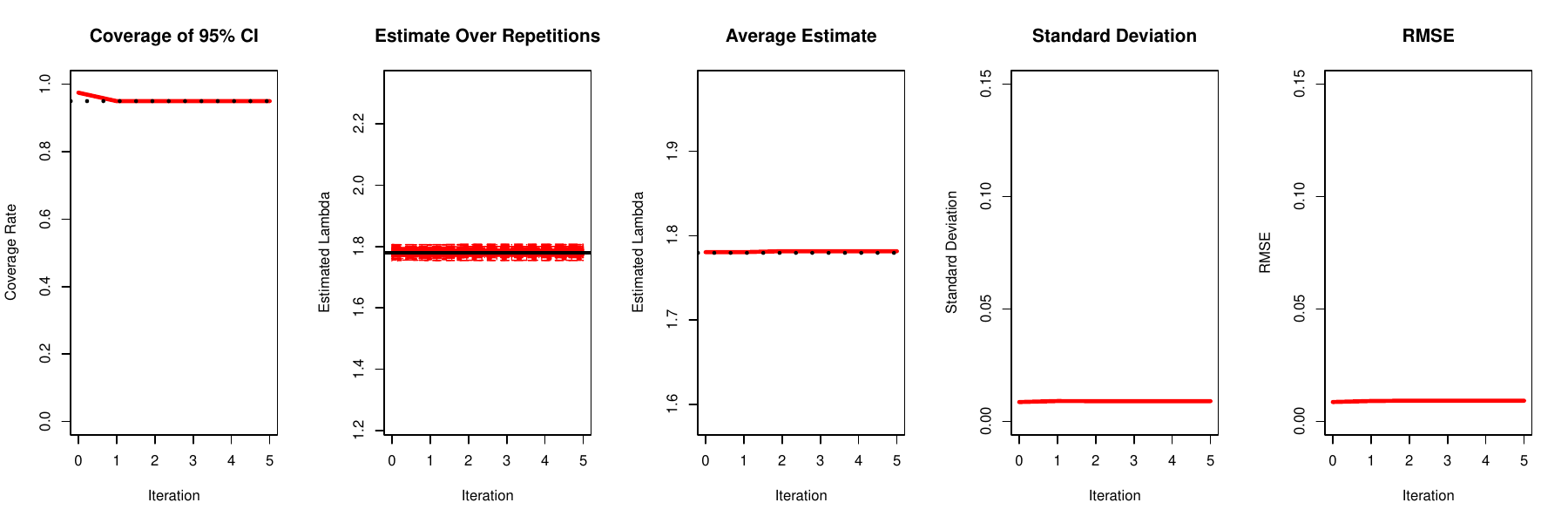}
    \caption{Results of the TMLE iterations for the target of inference $\lambda$ when the Rotifer data is considered. We consider 5 iterations, with 0 representing the initial estimate. We present results for the coverage of the 95\% confidence interval, estimates across repetitions, the average estimate, the standard deviation, and the root mean square error (RMSE). The simulation is repeated 200 times. }
    \label{fig:rotifer_lam}
\end{figure}

\begin{figure}[t]
    \centering
\includegraphics[width=\textwidth]{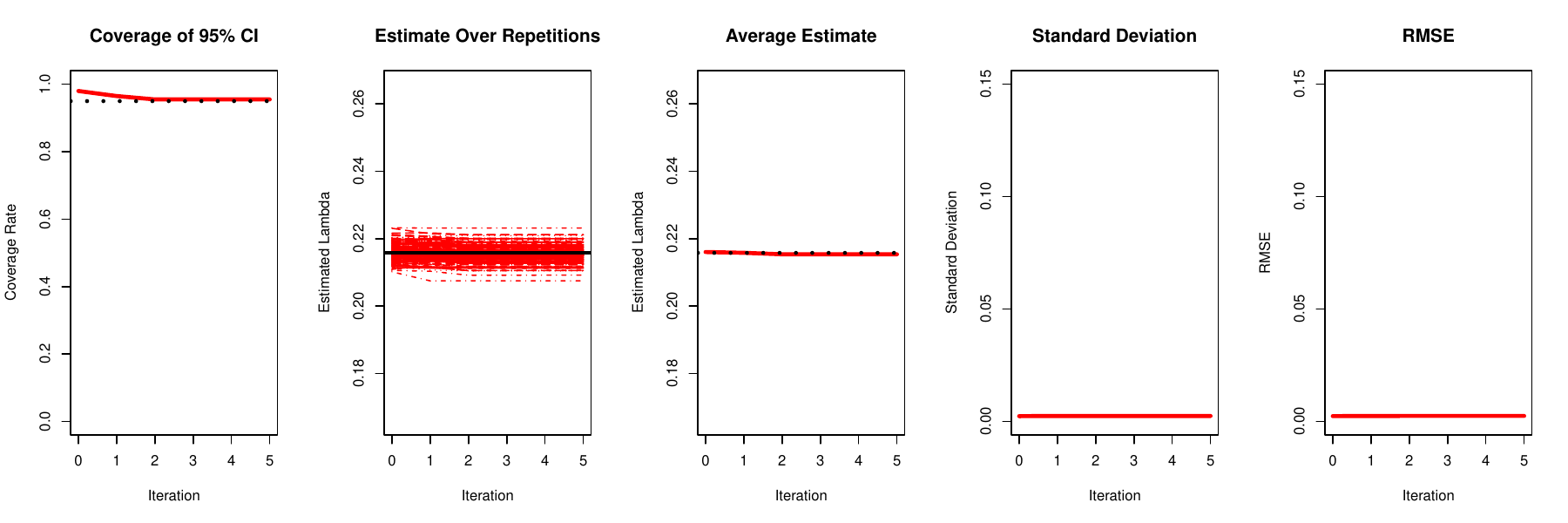}
    \caption{Results of the TMLE iterations for the target of elasticity when the Rotifer data is considered. We consider 5 iterations, with 0 representing the initial estimate. We present results for the coverage of the 95\% confidence interval, estimates across repetitions, the average estimate, the standard deviation, and the root mean square error (RMSE). The simulation is repeated 200 times. }
    \label{fig:rotifer_ela}
\end{figure}

Since the data is discrete and with no covariates, we can estimate the data distribution using the empirical distribution. This empirical distribution allows us to generate a synthetic dataset and serves as the initial estimate for the TMLE framework. The results are shown in Figures \ref{fig:rotifer_lam} and \ref{fig:rotifer_ela}. This experiment tests the hypothesis that if the initial estimate is already optimal (as the empirical distribution theoretically is), TMLE will not make further modifications.

\subsubsection{Real Data Analysis}
We conduct experiments using real data. Similar to the analysis with Idaho data, we employ the cross-fitting technique by splitting the dataset into five folds. The initial estimate is constructed using the empirical distribution of survival and fecundity. The estimates for both targets of inference, before and after the TMLE correction, along with the estimated confidence intervals, are presented in Figure \ref{fig:ci_real_rotifer}. We observe that TMLE does not make any significant corrections to the initial estimate, which is already optimal.

\section{Discussions} \label{sec:dis}
In this paper, we develop a robust and efficient estimator for key parameters in IPMs using the TMLE technique. Mathematically, we derive the efficient influence function for this specific type of target and construct the submodel path for the TMLE procedure. Our experiments include both simulations and real data analysis using Idaho and Rotifer data. Applying TMLE in practice is complex, primarily due to theoretical and numerical challenges. The derivation of the influence function is particularly demanding, making it difficult to apply our method to new targets of inference without careful mathematical calculations. This is more challenging when we also model the effect of time-varying individual covariates such as local competition; in such models our targets are often only calculable via simulating from the model output.  An alternative approach could involve using automatic differentiation to avoid repetitive math calculations. Numerically, we observe instability in TMLE when size classes are not well-balanced. We propose several solutions and suggest that future work could further explore semi-parametric models. We finally note that our methods remain reliant on assuming i.i.d. data and accounting for random effects or time-dependence remains a considerable future challenge.

\section*{Acknowledgements}

The authors gratefully acknowledge Stephen Ellner, Peter Adler, Robin Snyder and Christina Hernand\'{e}s for helpful discussions of the method and data, and for careful and constructive editing of the mansucript.  This work was supported by NSF grant DEB-1933497.

\bibliographystyle{imsart-nameyear} 
\bibliography{main}       

\appendix
\section{A Table of Mathematical Notation} \label{appendix:math_notations}

For the sake of clarity, we have provide a table of the notation we have used throughout the paper.

\begin{table}[H] \label{math:table}
\centering
\caption{Mathematical Notations Table}
\renewcommand{\arraystretch}{1.2} 
\setlength{\tabcolsep}{8pt} 
\begin{tabular}{|c|p{10cm}|}
\hline
\textbf{Symbol} & \textbf{Meaning} \\ \hline
$n(z,t)$        & Size distribution of individuals at time $t$ \\ \hline
$s(z)$          & Survival probability at size $z$ \\ \hline
$G(z', z)$      & Size transition kernel from size $z$ at time $t$ to size $z'$ at time $t+1$ \\ \hline
$F(z', z)$      & Per-capita production of new recruits of size z' by individuals of size $z$ \\ \hline
$K(z', z)$      & Net result of survival and reproduction: $K(z', z) = s(z)G(z', z) + F(z', z)$ \\ \hline
$Z_t$ & Size class of an individual at time $t$  \\ \hline
$S$ & The survival of
an individual beyond time $t$ \\ \hline
$Z^*_{t+1}$ &  The size class of an individual at time $t+1$ \\ \hline
$Y_j$ & The total number of offspring
falling into the size class $j$ \\ \hline
$\lambda$       & Long-term population growth (dominant eigenvalue) \\ \hline
$e$             & Elasticity of $\lambda$ with respect to fecundity \\ \hline
$\log \lambda_S$       & Average long-term population growth rate\\ \hline
$M$           & Survival matrix: a diagonal matrix representing survival. The subscript $\theta(t)$ is added when environmental covariates are taken into account.\\ \hline
$G$           & Growth matrix: transition probabilities between size classes. The subscript $\theta(t)$ is added when environmental covariates are taken into account. \\ \hline
$F$           & Fecundity matrix: expected number of offspring per size class. The subscript $\theta(t)$ is added when environmental covariates are taken into account.\\ \hline
$K$           & Combined kernel matrix: $K = M G + F$ \\ \hline
$u, v$      & Left and right dominant eigenvectors of the projection matrix $K$ \\ \hline
$P$ & Unknown data distribution \\ \hline
$P_n$           & Empirical distribution based on a sample of size $n$ \\ \hline
$\hat{P}$ &  Estimate of $P$ by using parametric or nonparametric model\\ \hline
$P_{\epsilon}$  & Perturbed distribution: $P_{\epsilon} = (1-\epsilon)P + \epsilon \hat{P}$ \\ \hline
$\epsilon$      & Small perturbation parameter used in TMLE submodel paths \\ \hline
$\Psi(P)$       & Functional mapping from distribution $P$ to the target of inference \\ \hline
$\psi(x;P)$     & Influence function describing the impact of a small perturbation in the distribution $P$ on the parameter $\Psi$ \\ \hline
$\tilde{\mathbb P}_n$ & Empirical distribution of $n$ samples drawn from the perturbed distribution $\tilde P_n$ \\ \hline
$\hat{p}$       & Initial estimate of the probability distribution \\ \hline
$\hat{\epsilon}$ & Optimal adjustment parameter in TMLE \\ \hline
$\hat{p}_{\epsilon}(x)$ & Adjusted probability distribution after TMLE update \\ \hline
$ C(\epsilon)$ & Normalizing term for constructing the submodel path $\hat{p}_{\epsilon}(x)$ \\ \hline
$\delta_x$      & Point mass distribution at $x$ \\ \hline
$R_2$           & Second-order remainder term in Taylor expansion \\ \hline
$V$ & Number of folds for CV-TMLE \\ \hline
$B_n$ & Cross-validation scheme \\ \hline
$\mathcal{T}$ & 
Index of the training sample \\ \hline
$\mathcal{V}$ & 
Index of the validation sample \\ \hline
$P_{B_n, n}^\mathcal{T}$ & Empirical probability distributions of the training samples \\ \hline
$P_{B_n, n}^\mathcal{V}$ & Empirical probability distributions of the validation samples \\ \hline
$\mathcal{I}_i$ & Index set for each fold \\ \hline
\end{tabular}
\end{table}

\section{Average long-term growth rate in relation to environmental covariates.} 
\label{appendix:lam}
Here we give a summary of some relevant theory about the long-term population growth rate in stochastic IPMs and MPMs of the form 
\begin{equation}
n(t+1) = K(\theta(t))n(t)
\end{equation}
based on \cite{ellner2007stochastic}. Here $K$ is the operator defined by the projection matrix or kernel, $n(t)$ is the population state (vector or distribution function) in year $t$, and $\theta(t)$ is the random environment (or environmental covariate) in year $t$. We assume that $\theta(t)$ is a stationary, ergodic stochastic
process. The population structure 
\begin{equation}
x(t)=n(t)/||n(t)||_1 
\end{equation}
then satisfies 
\begin{equation}
x(t+1) = \frac{K(\theta(t))x(t)}{||K(\theta(t))x(t)||_1}. 
\end{equation}
In the absence of environmental variation (so that the matrix/kernel is time-invariant), if the
matrix/kernel is power-positive, then (by the Perron-Frobenius Theorem) $x(t)$ converges to $u$, 
the normalized dominant right eigenvector of the matrix/kernel, for any initial population state. 
The dominant left eigenvector $v$ gives the state-dependent relative reproductive value of individuals. 
With environmental variation, under a similar positivity assumption (specifically, that for some $m$
the composition of $m$ sequential kernels is uniformly positive with probability 1), $x(t)$ does not
converge but it becomes asymptotically independent of $x(0)$ --- its value only depends (with vanishing
error as $t \to \infty$) on the sequence of environment states $\theta$ at times $0, 1, \cdots, t$. 
This behavior is what leads to properties B and C below. 

Note that the expected population $\bar n(t) = E n(t)$ satisfies $\bar{n}(t+1) = \bar{K} \bar{n}(t)$ 
where $\bar{K}$ is the mean matrix/kernel. So as $t \to \infty$, $\bar{n}(t)$ becomes proportional
to $u$, the dominant right eigenvector of the mean matrix/kernel. As dividing $\bar{n}$ by its norm
does not change the relative numbers in each state, $E x(t)$ converges to $u$. 

Under the positivity and ergodicity assumption stated above and mild technical assumptions (see \cite{ellner2007stochastic} for details) the following properties hold. 
\begin{itemize} 
\item[A.] The logarithm of the total population size $N(t)= ||n(t)||_1$ has a long-term growth rate $\log \lambda_S$ which is constant with probability 1,
\begin{equation}
\log \lambda_S =  
\mathop {\lim }\limits_{t \to \infty } t^{-1} \log N(t) = \mathop {\lim }\limits_{t \to \infty } t^{-1} E\log N(t). 
\end{equation}
The same is true for any weighted measure of population size $N_h(t)= \left\langle h,n(t) \right\rangle$ 
where $h \ge 0$ is a nonzero bounded measurable function. 

\smallskip 

\item[B.] Starting from any nonzero initial population state $n(z,0)=n_0(z)$ the population structure
$x(t)$ converges to a stationary random sequence of structures $\hat x(t)$ that is independent of 
$n_0$, and $E \hat{x}(t) = u$. 

\smallskip 

\item[C.] The joint sequence of environment states and stationary population 
structures $$( [\theta(0), \hat x(0)],[\theta(1), \hat x(1)], [\theta(2),\hat x(2)], \cdots )$$ 
is a stationary ergodic stochastic process. 

\smallskip 

\item[D.] The long-term growth rate can be computed as the average one-step growth rate, 
\begin{equation}
\log \lambda_S = E \log ||K(\theta(t))\hat x(t)||_1 
\label{eqn:lambda_t}
\end{equation} where the expectation is with respect to the joint stationary distribution 
in the previous item. 
\end{itemize}

On the right-hand side of \eqref{eqn:lambda_t},  $\hat \lambda(t) \equiv ||K(\theta(t))\hat x(t)||_1$ is the observed population growth rate from year $t$ to year $t+1$, once the population has converged onto the invariant joint measure of environment and population states. This is sometimes summarized by saying that 
\begin{equation}
\log \lambda_S = E \log \hat \lambda(t).
\label{eqn:lambdahat}
\end{equation}

However, $\hat \lambda(t)$ is not computable as the dominant eigenvalue of any kernel/matrix, or from the
dominant eigenvalues of $K(\theta)$ for different values of $\theta$. We therefore consider here the
the small-fluctuations approximation for $\lambda_S$, originally derived by Tuljapurkar 
\citep{tuljapurkar-1990}. Let $v$ and $u$ denote the dominant left and right
eigenvectors of the average kernel $\bar{K} = E K(\theta(t))$. Then to second order in the standard
deviation of the covariate (treated as a small parameter in a Taylor expansion),
\begin{equation}
\log \lambda_S \approx E \log \left( \frac{ \langle v, K(\theta(t))u \rangle}{\langle v, u \rangle} \right).
\label{eqn:SFA1}
\end{equation}

\section{Data Process of Idaho Data}\label{appendix:idaho_proc}
For each shrub species, individual-level datasets for each year were utilized, including multiple documents on survival, growth, and fecundity. Specifically, we summarize the data process procedure below:

\textbf{Growth and Survival Data:} The \texttt{growDnoNA.csv} file records the area size of each individual per year, while \texttt{survD.csv} provides information on whether an individual survived each year. These two tables were merged to obtain both survival and growth data for each individual. \textbf{Size Transformation and Discretization}
The area sizes were first log-transformed and then rescaled to a [0,1] range to create size values. These values were then discretized into size classes by creating 100 bins based on their percentiles.

\textbf{Fecundity Data:}
Fecundity data were obtained from \texttt{recArea.csv}, which records the total number of offspring produced in each quadrat per year, and \texttt{recSize.csv}, which contains the size values of these offspring. To incorporate recruitment into the IPM structure, we assumed that an individual's fecundity increases linearly with the square root of its area, akin to how a plant's seed production might grow with its size. Assuming a quadrat contains 
$\mathcal{J}$ non-seedling individuals, each with area size $a_i$ for $i \in \{1,2,\cdots, \mathcal{J}\}$, each offspring generated in the quadrat is assigned to a non-seedling individual $i$ as the parent with a probability $p_i = \frac{\sqrt{a_i}}{\sum_{i=1}^J \sqrt{a_i}}$. 

\textbf{Covariate Data:}
The \texttt{Climate.csv} file provides data on \texttt{ppt1}, \texttt{ppt2} (total precipitation over the past two years), and \texttt{TmeanSpr1}, \texttt{TmeanSpr2} (mean spring temperature over the past two years), which were used as baseline covariates. Additionally, \texttt{totParea} (total area currently covered by the species) was included as a covariate. We treat the year as a categorical factor and include it in the baseline covariates.

\section{Synthetic Data Generation of Idaho Data} \label{appendix:syn_idaho}
For the synthetic data analysis, we fit the machine learning model to the real data and use the fitted model to produce the synthetic samples. Here are the details for the technical procedures:

\textbf{Feature Generation: } For the distribution of the feature variables, we first fit a mixture distribution to $Z^c_t$:
\begin{align}
Z^c_t \sim w I(Z^c_t = 0) + (1-w) 
\text{Beta}(\beta_1,\beta_2)
\end{align}
This is done by initially estimating $w$ based on the proportion of  $Z^c_t$ that has a zero value (i.e., the seedlings). For the non-seedlings, we fit a beta distribution. The adaptive grid is then calculated according to the fitted distribution. We use a similar approach to fit the distribution of the Idaho distance weights variable. For the remaining covariates, we use the bootstrap method for sampling.

\textbf{Generation of $Z^{c}_{t+1}$: }
For the growth, we begin by fitting a linear regression with L1 norm regularization:
\begin{align}
Z^{c}_{t+1} \sim Z^c_t + \text{factor(year)} +  \text{ppt1}  +  \text{ppt2}  +  \text{TmeanSpr1}  +  \text{TmeanSpr2} +  \text{pptLag} + \epsilon
\end{align}
 For the residuals $\epsilon$ of the linear regression, we rescale them to a [0,1] range and fit them with a beta distribution. This procedure demonstrates generating $Z^c_{t+1}$ by sampling from the beta distribution, rescaling the values, and then adding the prediction from the fitted linear model.

\textbf{Generation of $S$: }
For the survival, we consider two separate logistc regressions with L1 norm regularization, one for seedlings and another for non-seedlings:
\begin{align}
S_{\text{seedling}} &\sim \text{logit}\Big (  \text{factor(year)} +  \text{ppt1}  +  \text{ppt2}  +  \text{TmeanSpr1}  +  \text{TmeanSpr2} +  \text{pptLag}\Big ) \\
S_{\text{nonseedling}} &\sim \text{logit}\Big ( Z^c_t  + \text{factor(year)} +  \text{ppt1}  +  \text{ppt2}  +  \text{TmeanSpr1}  +  \text{TmeanSpr2} +  \text{pptLag}\Big ) 
\end{align}
The $S$ can be simulated using a Bernoulli distribution, with the probability given by the output of the logistic regression model for seedling or nonseedling.

\textbf{Generation of $Y_j$: }
For the fecundity, we begin with the data of quadrat level. We fit a Poisson regression with L1 norm regularization:
\begin{align}
Y \sim \text{logit}\Big ( \text{factor(year)} +  \text{ppt1}  +  \text{ppt2}  +  \text{TmeanSpr1}  +  \text{TmeanSpr2} +  \text{pptLag}\Big ) 
\end{align}
where $Y$ is the total number of offsprings at each quadrat.  Then assuming a quadrat contains $\mathcal{J}$ non-seedling individuals, each with area size $a_i$ for $i \in \{1,2,\cdots, \mathcal{J}\}$, each offspring generated in the quadrat is assigned to a non-seedling individual $i$ as the parent with a probability $p_i = \frac{\sqrt{a_i}}{\sum_{i=1}^J \sqrt{a_i}}$. 
Each offspring is then randomly assigned to a size class based on the distribution of size classes within the total population of recruits.

\section{Modeling Process of Idaho Data} \label{appendix:model_idaho}

\textbf{Growth Model: }
For the growth model, we first fit a linear regression with L1 norm regularization:
\begin{align}
Z^{c}_{t+1} \sim Z^c_t + \text{factor(year)} +  \text{ppt1}  +  \text{ppt2}  +  \text{TmeanSpr1}  +  \text{TmeanSpr2} +  \text{pptLag} + \epsilon
\end{align}
 For the residuals $\epsilon$ of the linear regression, we rescale them to a [0,1] range. Then we consider kernel density estimation with different bandwidths
(0.01, 0.02, 0.03, 0.05, 0.01) and also consider the case where cross-validation is employed to select the optimal bandwidth.  

\textbf{Survival Model: }
For the survival, we consider two separate logistc regressions with L1 norm regularization, one for seedlings and another for non-seedlings:
\begin{align}
S_{\text{seedling}} &\sim \text{logit}\Big (  \text{factor(year)} +  \text{ppt1}  +  \text{ppt2}  +  \text{TmeanSpr1}  +  \text{TmeanSpr2} +  \text{pptLag}\Big ) \\
S_{\text{nonseedling}} &\sim \text{logit}\Big ( Z^c_t  + \text{factor(year)} +  \text{ppt1}  +  \text{ppt2}  +  \text{TmeanSpr1}  +  \text{TmeanSpr2} +  \text{pptLag}\Big ) 
\end{align}

\textbf{Fecundity: }
For the fecundity, we consider two distinct Poisson regression models with L1 norm regularization, predicting the expected number of offspring that become seedlings or non-seedlings:
\begin{align}
Y_{\text{seedling}} &\sim \text{logit}\Big (Z^c_t  + \text{factor(year)} +  \text{ppt1}  +  \text{ppt2}  +  \text{TmeanSpr1}  +  \text{TmeanSpr2} +  \text{pptLag}\Big ) \\
Y_{\text{nonseedling}} &\sim \text{logit}\Big (Z^c_t + Z_r  + \text{factor(year)} +  \text{ppt1}  +  \text{ppt2}  +  \text{TmeanSpr1}  +  \text{TmeanSpr2} +  \text{pptLag}\Big )
\end{align}
where $Z_r$ represents the size value corresponding to the size class into which the offspring is categorized.

\section{Modeling of Rotifer Data} \label{appendix_rotifer_process}
There are 16 age classes in total. And we have four classes for the maternal age in the dataset, that are 3, 5, 7 and 9 days old. For simplicity, we construct four maternal age groups for the offsprings:
\begin{itemize}
    \item Maternal age group 1: the offsprings that are generated by the parents with age 1, 2, 3, or 4 d.
    \item Maternal age group 2: the offsprings that are generated by the parents with age 5 or 6 d.
    \item Maternal age group 3: the offsprings that are generated by the parents with age 7 or 8 d.
    \item Maternal age group 4: the offsprings that are generated by the parents with age 9-16 d.
\end{itemize}

We denote $n_{i, j}(t)$ is the number of individuals in maternal age group $i$  ($i=1,2,3,4$) and age class $j$ ($j=1,2,\cdots,16$) at time $t$. Then the composition of the population is given by a column vector $\tilde{\mathbf{n}}(t)$ that collects maternal ages within age classes:
$$\tilde{\mathbf{n}}(t)=\left(\begin{array}{c}n_{1,1}(t) \\ \vdots \\ n_{1, 16}(t) \\ \hline \vdots \\ \hline n_{4, 1}(t) \\ \vdots \\ n_{4, 16}(t)\end{array}\right)$$

An individual with maternal age $i$ and age $j$ produces $f_{i j}$ daughters in $1 \mathrm{~d}$ and survives to age $j+1$ with probability $p_{i j}$. These vital rates are incorporated into a fertility matrix $\tilde{\mathbf{F}}$ and a survival matrix $\tilde{\mathbf{U}}$. The population projection matrix $\tilde{\mathbf{A}}$, which projects the population vector from one day to the next, is the sum of $\tilde{\mathbf{F}}$ and $\tilde{\mathbf{U}}$, and the population dynamics are given by
$$
\tilde{\mathbf{n}}(t+1)=(\tilde{\mathbf{U}}+\tilde{\mathbf{F}}) \tilde{\mathbf{n}}(t)=\tilde{\mathbf{A}} \tilde{\mathbf{n}}(t) .
$$

The Matrix Ū. This block matrix takes the form

$$
\tilde{\mathbf{U}}=\left(\begin{array}{cccc}
\mathbf{U}_1 & 0 &  0 & 0  \\
0 & \mathbf{U}_2 & 0 & 0 \\
 0 & 0 & \mathbf{U}_3  & 0\\
 0 & 0 & 0  & \mathbf{U}_{4}
\end{array}\right) .
$$

The $\mathbf{U}_j$ are $16 \times 16$ matrices with survival probabilities on the diagonal and zeros elsewhere:
$$
\mathbf{U}_j=\left(\begin{array}{ccccc}
0 & & & &\\
p_{1, j} & 0 & & &\\
& p_{2, j} & 0 & &\\
& & \ddots & 0 &\\
& & & p_{15, j} & 0
\end{array}\right) .
$$

The Matrix $\tilde{\mathbf{F}}$ (Row 1 represents the offspring belongs to age class 1 and maternal age group 1. Row 17 represents the offspring belongs to age class 1 and maternal age group 2. Row 33 represents the offspring belongs to age class 1 and maternal age group 3. And row 49 represents the offspring belongs to age class 1 and maternal age group 4.):





$$
\tilde{\mathbf{F}}=\left(\begin{array}{cccc}
\mathbf{f}^{(1)}_{1} & \mathbf{f}^{(2)}_{1} & \mathbf{f}^{(3)}_{1} & \mathbf{f}^{(4)}_{1}  \\ 0 & 0 & 0 & 0\\ \cdots & \cdots & \cdots & \cdots \\ 0 & 0 & 0 & 0 \\ \mathbf{f}^{(1)}_{17} & \mathbf{f}^{(2)}_{17} & \mathbf{f}^{(3)}_{17} & \mathbf{f}^{(4)}_{17}  \\ 0 & 0 & 0 & 0\\ \cdots & \cdots & \cdots & \cdots \\ 0 & 0 & 0 & 0 \\ \mathbf{f}^{(1)}_{33} & \mathbf{f}^{(2)}_{33} & \mathbf{f}^{(3)}_{33} & \mathbf{f}^{(4)}_{33}  \\ 0 & 0 & 0 & 0\\ \cdots & \cdots & \cdots & \cdots \\ 0 & 0 & 0 & 0 \\ \mathbf{f}^{(1)}_{49} & \mathbf{f}^{(2)}_{49} & \mathbf{f}^{(3)}_{49} & \mathbf{f}^{(4)}_{49}  \\ 0 & 0 & 0 & 0\\ \cdots & \cdots & \cdots & \cdots \\ 0 & 0 & 0 & 0 \\
\end{array}\right), \quad \quad \tilde{\mathbf{n}}(t)=\left(\begin{array}{c} n_{1,1}(t) \\ \vdots \\ n_{1, 16}(t) \\ \hline n_{2,1}(t) \\ \vdots \\ n_{2, 16}(t) \\ \hline n_{3,1}(t) \\ \vdots \\ n_{3, 16}(t) \\ \hline n_{4, 1}(t) \\ \vdots \\ n_{4, 16}(t)\end{array}\right)
$$

where $$\mathbf{f}^{(j)}_{1}  = (f^{(j)}_{1,1},f^{(j)}_{1,2},f^{(j)}_{1,3},f^{(j)}_{1,4},0,\cdots,0)$$ $$\mathbf{f}^{(j)}_{17} = (0,0,0, f^{(j)}_{17,5},f^{(j)}_{17,6},0,\cdots,0)$$ $$\mathbf{f}^{(j)}_{33} = (0,0,0, 0, 0, 0, f^{(j)}_{33,7},f^{(j)}_{33,8},0,\cdots,0)$$ $$\mathbf{f}^{(j)}_{49} = (0,\cdots, 0, f^{(j)}_{49,9},f^{(j)}_{49,10},f^{(j)}_{49,11},f^{(j)}_{49,12},f^{(j)}_{49,13},f^{(j)}_{49,14},f^{(j)}_{49,15},f^{(j)}_{49,16})$$

For example, if we multiply the first row of $\tilde{\mathbf{F}}$ with $\tilde{\mathbf{n}}(t)$, we have $f^{(1)}_{1,1}n_{1,1}(t) + f^{(1)}_{1,2}n_{1,2}(t)+f^{(1)}_{1,3}n_{1,3} + f^{(1)}_{1,4} n_{1,4} + f^{(2)}_{1,1}n_{2,1}(t) + f^{(2)}_{1,2}n_{2,2}(t)+f^{(2)}_{1,3}n_{1,3} + f^{(12}_{1,4} n_{1,4} + f^{(3}_{1,1}n_{1,1}(t) + f^{(3)}_{1,2}n_{1,2}(t)+f^{(3)}_{1,3}n_{1,3} + f^{(3)}_{1,4} n_{1,4} + f^{(4)}_{1,1}n_{1,1}(t) + f^{(4)}_{1,2}n_{1,2}(t)+f^{(4)}_{1,3}n_{1,3} + f^{(4)}_{1,4} n_{1,4}$

\section{Additional Results for Elasticity}
\begin{figure}[H]
    \centering
    \includegraphics[width=\textwidth]{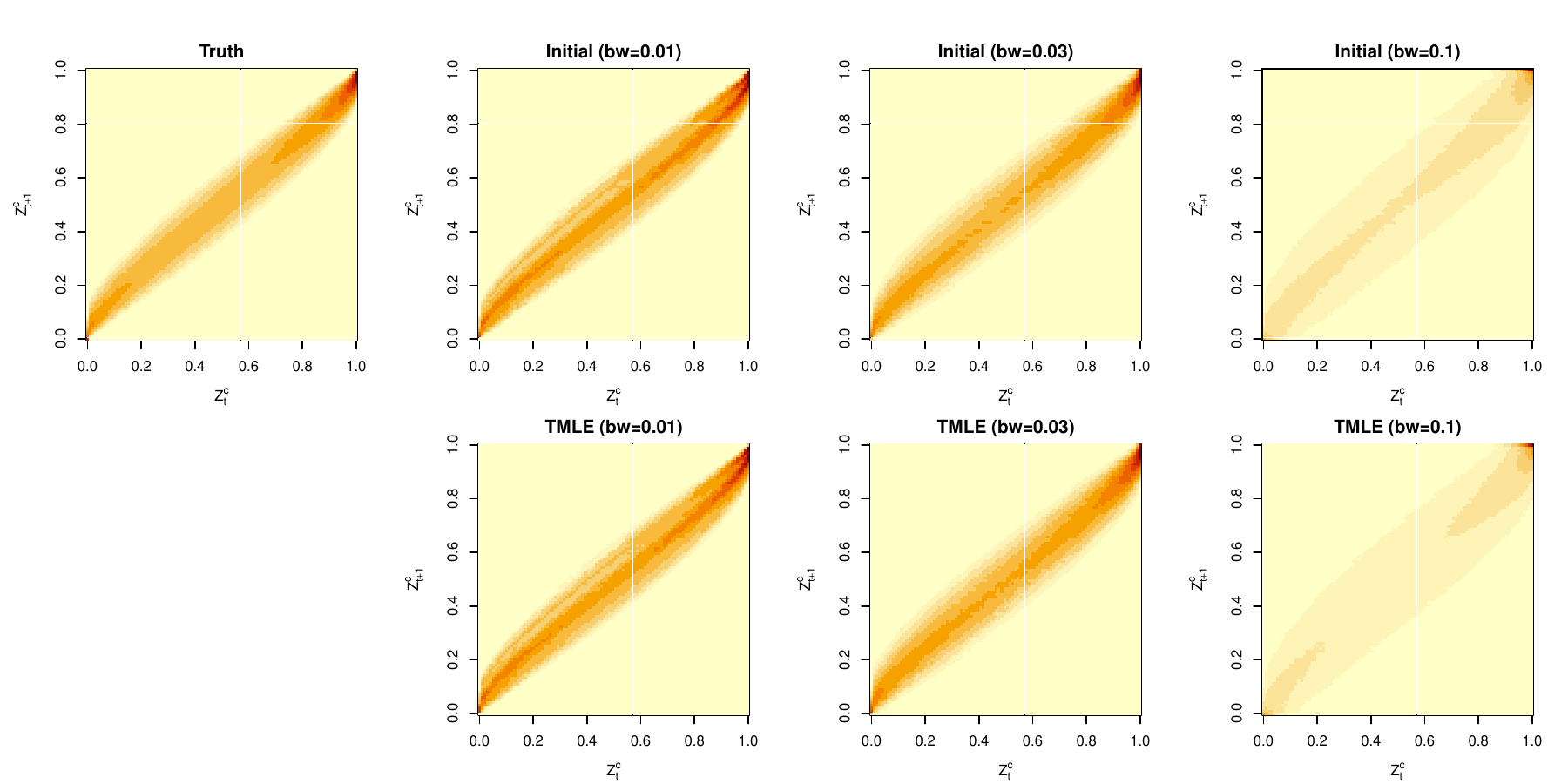}
    \caption{Heatmap for the matrix multiplication $MG$ when the target  is the elasticity. We use the true matrix as the baseline for comparison. The term 'Initial' refers to the initially estimated matrix without any bias correction by TMLE. 'TMLE' indicates the results after applying the TMLE update. We present results for different bandwidths (bw) of $0.01, 0.03$ and $0.1$. }
    \label{fig:sg_ela}
\end{figure}

\begin{figure}[H]
    \centering
\includegraphics[width=\textwidth]{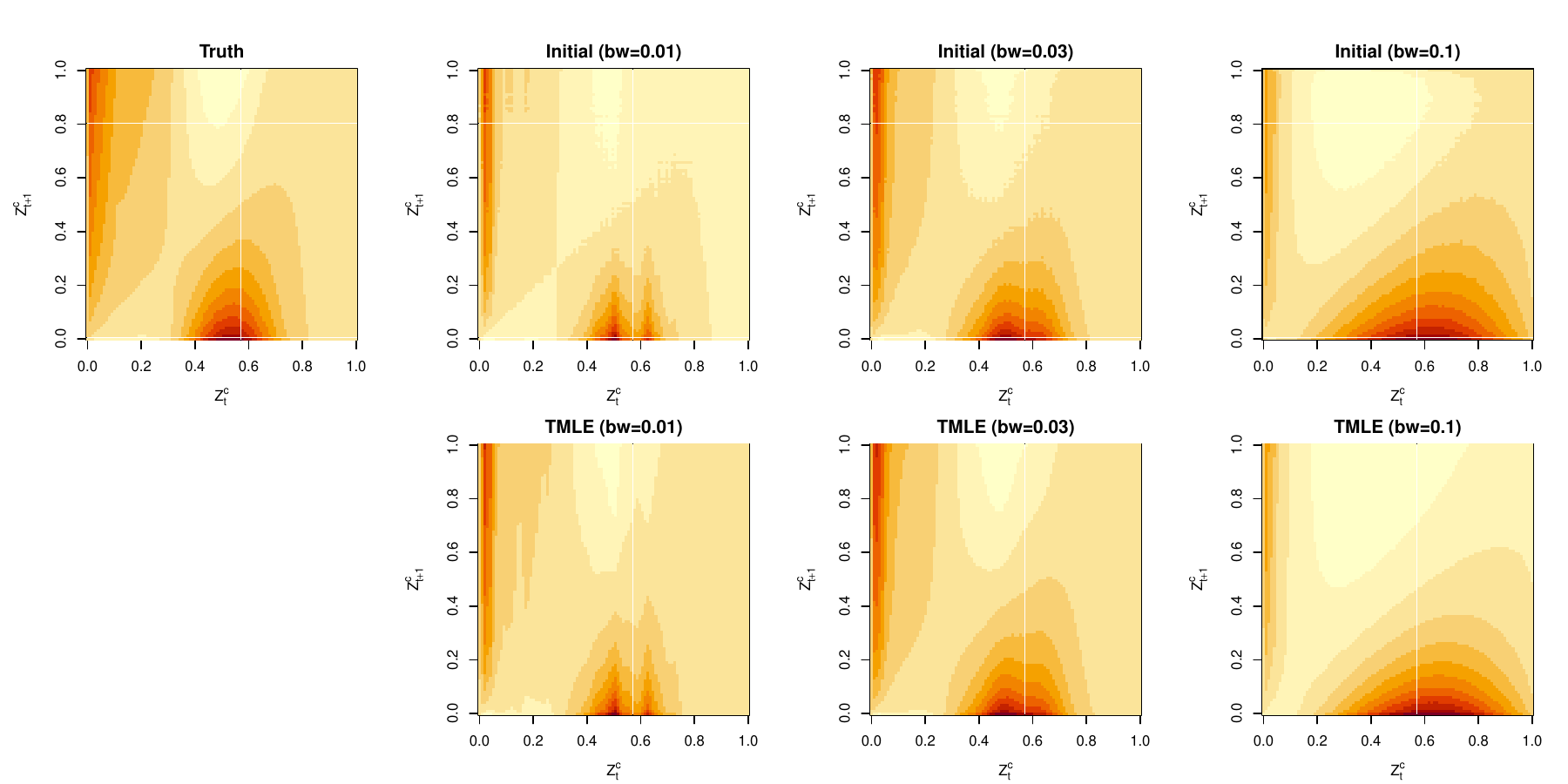}
    \caption{Heatmap for the efficient influence function of $MG$ when the target is the elasticity. We use the true matrix as the baseline for comparison. The term 'Initial' refers to the initially estimated matrix without any bias correction by TMLE. 'TMLE' indicates the results after applying the TMLE update. We present results for different bandwidths (bw) of $0.01, 0.03$ and $0.1$. }
    \label{fig:eig_ela}
\end{figure}

\begin{figure}[H]
    \centering
\includegraphics[width=\textwidth]{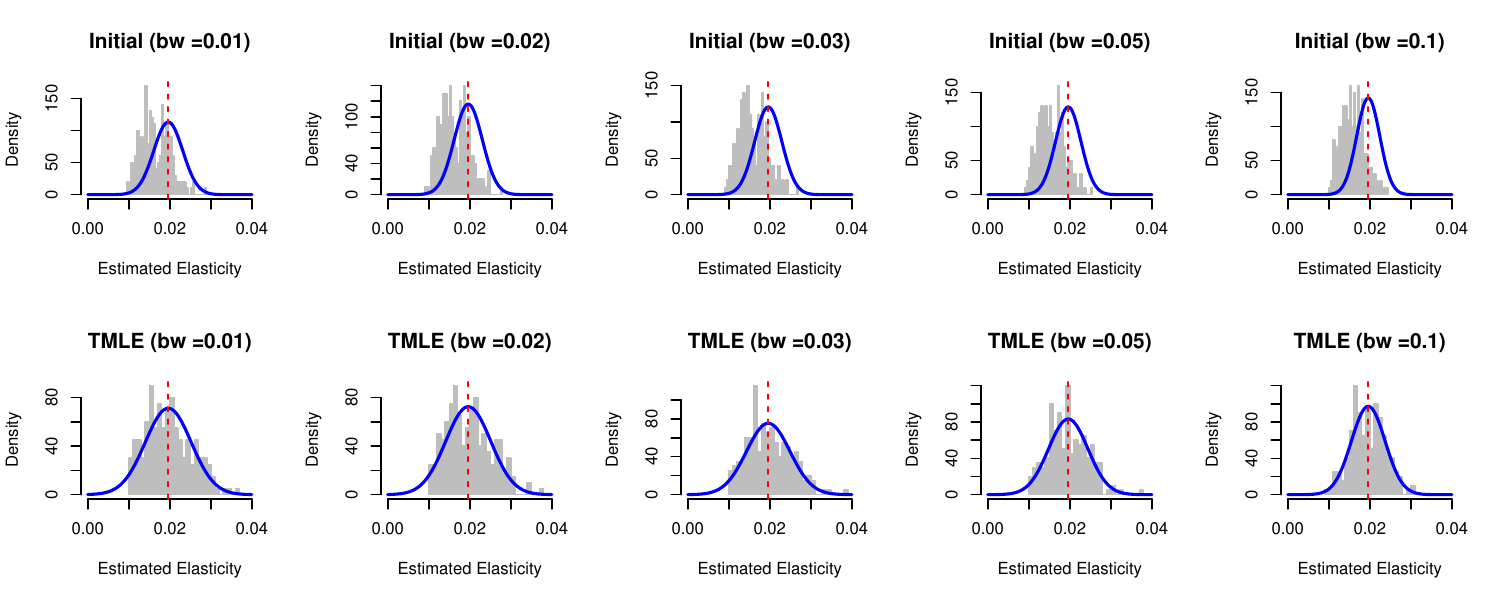}
    \caption{Histogram for the estimated elasticity across 200 repetitions. The term 'Initial' refers to the initially estimated matrix without any bias correction by TMLE. 'TMLE' indicates the results after applying the TMLE update. We present results for different bandwidths (bw) of $0.01, 0.02, 0.03, 0.05$ and $0.1$.  The blue curve represents the density of a normal distribution with the true mean and the standard deviation of the estimators.}
    \label{fig:den_ela}
\end{figure}

\begin{figure}[H]
    \centering
\includegraphics[width=\textwidth]{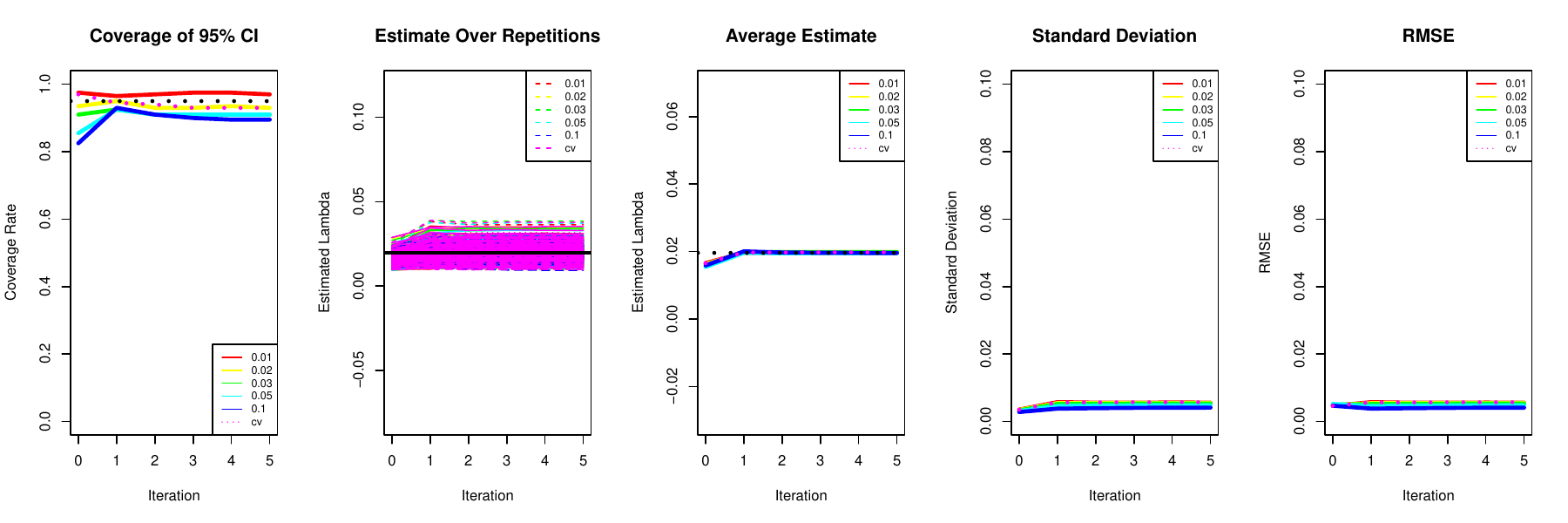}
    \caption{Results of the TMLE iterations for the target elasticity. We consider 5 iterations, with 0 representing the initial estimate. Different bandwidths (bw)  of $0.01, 0.02, 0.03, 0.05$ and $0.1$are evaluated, including the optimal bandwidth selected by cross-validation (cv). We present results for the coverage of the 95\% confidence interval, estimates across repetitions, the average estimate, the standard deviation, and the root mean square error (RMSE). The simulation is repeated 200 times. }
    \label{fig:ela}
\end{figure}

\section{Derivations of Influence Functions} \label{appendix:eif}
\subsection{Proof of Theorem \ref{thm1}}
Here we use the tricks of point mass contamination. We perturb the true distribution by
\begin{align}
P_{\epsilon} (Z^*_{t+1}=z^*_{t+1}|Z_t = z_t) &=  \frac{P_{\epsilon} \Big((Z^*_{t+1},Z_{t})=(z^*_{t+1},z_t)\Big)}{P_{\epsilon} (Z_{t}=z_t)} \\
&= \frac{(1-\epsilon)P \Big((Z^*_{t+1},Z_{t})=(z^*_{t+1},z_t)\Big) + \epsilon 1_{(z'_t,z'_{t+1})}}{(1-\epsilon)P (Z_{t}=z_t)+\epsilon 1_{z'_t}}
\end{align}
\begin{align}
P_{\epsilon} (Y_{j}=y_{j}|Z_t = z_t) &=  \frac{P_{\epsilon} \Big((Y_{j},Z_{t},X)=(y_{j},z_t)\Big)}{P_{\epsilon} (Z_{t}=z_t)} \\
&= \frac{(1-\epsilon)P \Big((Y_{j},Z_{t})=(y_{j},z_t)\Big) + \epsilon 1_{(y'_{j},z'_t)}}{(1-\epsilon)P (Z_{t}=z_t)+\epsilon 1_{(z'_t)}}
\end{align}
where $1_x$ is a point mass distribution with all the probability at $x$.

We denote $(GM)_{\epsilon}$ as the matrix for perturbed distribution, that is $\{(GM)_{\epsilon}\}_{ji} = P_{\epsilon}(Z^*_{t+1} = j|Z_t = i)$ and define $F_{\epsilon}$ by

\begin{align}
\{F_{\epsilon}\}_{ji} =  \sum_{y_j} y_j \frac{(1-\epsilon)P \Big((Y_j,Z_{t})=(y_j,i)\Big) + \epsilon 1_{(y'_{t+1},z'_t)}}{(1-\epsilon)P (Z_{t}=i)+\epsilon 1_{z'_t}}
\end{align}

Then we have
\begin{align}
\begin{split}
\frac{d\Psi(P_{\epsilon})}{d\epsilon} \Big |_{\epsilon=0} &= \frac{d \max\text{eigen}((GM)_{\epsilon}+F_{\epsilon})}{d\epsilon} \Big |_{\epsilon=0} \\  
&=v^T \frac{d((GM)_{\epsilon}+F_{\epsilon})}{d\epsilon}\Big |_{\epsilon=0} u
\end{split}
\end{align}
since $a$ is a constant. Specifically, for each $i,j$, we have that 
\begin{align}
&\frac{d\{(GM)_{\epsilon}\}_{ji}}{d\epsilon}\Big |_{\epsilon=0} \\
&= \frac{dP_{\epsilon} (Z^*_{t+1}=j|Z_t = i)}{d\epsilon}\Big |_{\epsilon=0} \\
&= \scalebox{0.95}{$\frac{\left[1_{(z'_t,z'_{t+1})} - P\Big((Z_t,Z^*_{t+1})=(i,j)\Big)\right]P(Z_t=i) - P\Big((Z_t,Z^*_{t+1})=(i,j)\Big)\Big[1_{z'_t} - P(Z_t=i)\Big]}{P^2(Z_t=i)}$}\\
&= \scalebox{0.95}{$\frac{1_{(z'_t,z'_{t+1})} - P\Big((Z_t,Z^*_{t+1})=(i,j)\Big)}{P(Z_t=i)} - \frac{ 1_{z'_t} - P(Z_t=i)}{P(Z_t=i)}P\Big(Z^*_{t+1}=j|Z_t=i\Big)$} \\
&=  \frac{1_{(z'_t,z'_{t+1})} - 1_{z'_t} P(Z^*_{t+1}=j|Z_t=i)}{P(Z_t=i)} \\
\end{align}
Similarly,
\begin{align}
\frac{d\{F_{\epsilon}\}_{ji}}{d\epsilon}\Big |_{\epsilon=0} =  \frac{y'_{j}1_{z'_t} - 1_{z'_t} E(Y_j|Z_t=i)}{P(Z_t=i)} 
\end{align}
Plug everything into (1) and with point mass contamination, we get that the efficient influence function is 
\begin{align}
 \phi(Z_{t},Z^*_{t+1},Y_j) = u^T(W_{1}+W_{2})v
\end{align}
where
\begin{align}
W_{1,ji} =\frac{I\{Z_t=i,Z^*_{t+1}=j\} - I\{Z_t=i\} P(Z^*_{t+1}=j|Z_t=i)}{P(Z_t=i)} 
\end{align}
and
\begin{align}
W_{2,ji} =\frac{I\{Z_t=i\}\Big(Y_j- E(Y_j|Z_t=i)\Big)}{P(Z_t=i)} 
\end{align}

Since $P(Z_t,Z^*_{t+1},Y_1,\cdots,Y_d) = P(Z_t,Z^*_{t+1})P(Y_1,\cdots,Y_d|Z_t,Z^*_{t+1})$,  we decompose the whole model space into  submodel spaces for growth and survival $\$Z^*_{t+1}|Z_t$ and reproduction $\{Y_1,\cdots,Y_d|Z_t,Z^*_{t+1}\}$. In this paper, we make the independence assumption that $\{Y_1,\cdots,Y_d\} \perp Z^*_{t+1} | Z_{t}$. Therefore, the submodel space $\{Y_1,\cdots,Y_d|Z_t\}$ can be equivalently written into $\{Y_1,\cdots,Y_d|Z_t\}$. We project the EIF into each corresponded subtangent space. For the submodel space $Z^*_{t+1}|Z_t$,
\begin{align}
\phi_{Z^*_{t+1}|Z_t}(Z_{t},Z^*_{t+1}) &= E\Big[\phi(Z_{t},Z^*_{t+1},Y_1,\cdots,Y_d)|Z_t,Z^*_{t+1} \Big ] \\
& = v^T W_{1} u
\end{align}
For the submodel space  $Y_1,\cdots,Y_d|Z_t$,
\begin{align}
& \phi_{Y_1,\cdots,Y_d|Z_t}(Z_{t},Z^*_{t+1},Y_1,\cdots,Y_d) \\
= \; &\phi(Z_{t},Z^*_{t+1},Y_1,\cdots,Y_N) - E\Big[\phi(Z_{t},Z^*_{t+1},Y_1,\cdots,Y_d)|Z_t\Big]   \\
=  \; & v^T W_{2} u
\end{align}

\subsection{Proof of Theorem \ref{thm2}}
We use the delta method and apply the chain rule:
\begin{align}
& \quad \frac{\partial e_{\epsilon}}{\partial \epsilon} \Big \vert_{\epsilon=0} \\ &= E \Big [\frac{\partial [\frac{v^T_{\epsilon} F_{\epsilon} u_{\epsilon}}{\lambda_{\epsilon}\langle v_{\epsilon}, u_{\epsilon} \rangle}]}{\partial \epsilon} \Big \vert_{\epsilon=0} \Big ] \\
&= E \Big [\frac{ A - B}{\lambda^2\langle  v,u\rangle^2}\Big ] 
\end{align}
where 
\begin{align}
A &:= \Big[\frac{\partial v^T_{\epsilon}}{\partial \epsilon} \vert_{\epsilon=0} F u + v^T \frac{\partial F_{\epsilon}}{\epsilon}\vert_{\epsilon=0} u +v^T F \frac{\partial u_{\epsilon}}{\partial \epsilon} \vert_{\epsilon=0}  \Big]\lambda\langle v,u \rangle \\
B &:= v^T F u \Big [ \frac{\partial \lambda_{\epsilon}}{\partial \epsilon} \langle v,u \rangle + \lambda \langle \frac{\partial v_{\epsilon}}{\partial \epsilon} \vert_{\epsilon=0},u\rangle 
 + \lambda \langle v, \frac{\partial u_{\epsilon}}{\partial \epsilon} \vert_{\epsilon=0}\rangle \Big]
\end{align}
\begin{align}
\frac{\partial v^T_{\epsilon}}{\partial \epsilon} \vert_{\epsilon=0} =  v^T\Big (\frac{\partial K_{\epsilon}}{\partial \epsilon}\vert_{\epsilon=0} -\frac{\partial \lambda_{\epsilon}}{\partial  \epsilon}\vert_{\epsilon=0}  I \Big )(\lambda I - K)^+, \end{align}
\begin{align}
\frac{\partial v_{\epsilon}}{\partial \epsilon} \vert_{\epsilon=0} = (\lambda I-K)^+ \Big (\frac{\partial K_{\epsilon}}{\partial \epsilon}\vert_{\epsilon=0} -\frac{\partial \lambda_{\epsilon}}{\partial  \epsilon}\vert_{\epsilon=0}  I \Big ) v 
\end{align}

We denote 
\begin{align}
W_{1,ij} =\frac{I\{Z_t=i,Z^*_{t+1}=j\} - I\{Z_t=i\} P(Z^*_{t+1}=j|Z_t=i)}{P(Z_t=i)} 
\end{align}
and
\begin{align}
W_{2,ij} =\frac{I\{Z_t=i\}\Big(Y_j- E(Y_j|Z_t=i)\Big)}{P(Z_t=i)} 
\end{align}
for $i,j = 1,2, \cdots, N$.



Then we introduce the notations
\begin{align}
v^T_{1} = v^T (\lambda_X I - K)^+ \qquad  v^T_{2} = v^TF (\lambda I - K)^+  \\
u_{1} =  (\lambda I - K)^+ u \qquad u_{2} =  (\lambda I - K)^+ F u  \\
 c = \frac{v^T F u}{\langle v,u \rangle } \quad
\widetilde{W}_{1} = W_{1} - \frac{v^T W_{1}u}{\langle v,u\rangle} \quad \widetilde{W}_{2} = W_{2} - \frac{v^T W_{2}u}{\langle v,u\rangle}
\end{align}

Then we have
\begin{align}
\phi(Z_{t},Z^*_{t+1},Y_1,\cdots,Y_N) = \frac{C -   D}{\lambda_X^2 \langle v,u \rangle}
\end{align}
where
\begin{align}
C &:=  \Big [ v^T (\widetilde{W}_{1}+\widetilde{W}_{2} ) u_{2} + v^TW_{2}u + v^T_{2} (\widetilde{W}_{1}+\widetilde{W}_{2} ) u\Big] \lambda   \\
D &:=  c\Big [ v^T (\widetilde{W}_{1}+\widetilde{W}_{2} )u \langle v,u\rangle + \lambda v^T (\widetilde{W}_{1}+\widetilde{W}_{2})u_{1} + \lambda v^T_{1} (\widetilde{W}_{1}+\widetilde{W}_{2})u]
\end{align}
Then we get

\begin{align}
&\psi_{\text{Elasticity};Z^*_{t+1}|Z_t}(Z_{t},Z^*_{t+1},Y_1,\cdots,Y_N) \\ & = \frac{ \Big [ v^T \widetilde{W}_{1} u_{2} + v^T_{2} \widetilde{W}_{1} u\Big] \lambda -  c  \Big [ v^T \widetilde{W}_{1}u \langle v,u\rangle + \lambda v^T \widetilde{W}_{1}u_{1} + \lambda v^T_{1} \widetilde{W}_{1}u]}{\lambda^2 \langle v,u \rangle} \\
&\psi_{\text{Elasticity};Y_1,\cdots,Y_N|Z_t}(Z_{t},Z^*_{t+1},Y_1,\cdots,Y_N) \\ & = \frac{ \Big [ v^T \widetilde{W}_{2} (u_{2}+u)  + v^T_{2} \widetilde{W}_{2} u\Big] \lambda -  c  \Big [ v^T \widetilde{W}_{2}u \langle v,u\rangle + \lambda v^T \widetilde{W}_{2}u_{1} + \lambda v^T_{1} \widetilde{W}_{2}u]}{\lambda^2 \langle v,u \rangle}
\end{align}

\subsection{Proof of Theorem}
Assume that we have the data sampled from $(Z_t,Z^*_{t+1},S,Y_1,\cdots$ $,Y_N,\theta(t))$, where $Z_t$ is the size class at time point $t$, $Z^*_{t+1}$ is the size class at time point $t+1$, with value $0$ representing the death. $S$ is the survival and $Y_j$ is the number of produced offsprings belong to the size class $j$. $\theta(t)$ is all the random environment in year $t$.

To construct a traditional estimate, we can use parametric models or kernel density estimation to calculate $P(Z^*_{t+1}=j|Z_t=i,S=1,\theta(t)=\theta$), $P(S=1|Z_t=i,\theta(t))$ and $E(Y_j|Z_t=i,\theta(t)=\theta)$. Then this gives the estimate of the matrices $M_{\theta(t)}G_{\theta(t)}$,  $F_{\theta(t)}$. Then $K_{\theta(t)} = G_{\theta(t)}M_{\theta(t)} + F_{\theta(t)}$. 

We now consider the target of inference
\begin{equation}
\log \lambda_S \approx E \log \left( \frac{v^TK_{\theta(t)}u}{\langle v,u \rangle} \right).
\end{equation}
where $v$ and $u$ represents the dominant left and right eigenvalues of the average matrix $\bar{K} = E K(\theta(t))$. The expectation is taken over the randomness of $\theta(t)$. We now perturb the distribution of the target  by an infinitesimal amount $\epsilon$, this gives
\begin{equation}
\log \lambda_{S,\epsilon} \approx E \log \left( \frac{v_{\epsilon}^TK_{\epsilon}(\theta(t))u_{\epsilon}}{\langle v_{\epsilon},u_{\epsilon}\rangle} \right).
\end{equation}

Taking the derivative at $\epsilon = 0$, it gives that
\begin{align}
\frac{\partial \log \lambda_{S,\epsilon}}{\partial \epsilon} \vert_{\epsilon=0}&= E\left( \frac{\langle v,u \rangle} {v^TK_{\theta(t)}u}  \frac{A-B}{\langle v,u\rangle^2}\right) \\
&= E\left( \frac{A-B} {v^TK_{\theta(t)}u \langle v,u\rangle}\right) 
\end{align}
where 
\begin{align*}
A &:= \Big[\frac{\partial v^T_{\epsilon}}{\partial \epsilon} \vert_{\epsilon=0} K_{\theta(t)} u + v^T \frac{\partial K_{\theta(t),\epsilon}}{\epsilon}\vert_{\epsilon=0} u +v^T K_{\theta(t)} \frac{\partial u_{\epsilon}}{\partial \epsilon} \vert_{\epsilon=0}  \Big]\langle v,u \rangle \\
B &:= v^T K_{\theta(t)} u \Big [ \langle \frac{\partial v_{\epsilon}}{\partial \epsilon} \vert_{\epsilon=0},u\rangle 
 +  \langle v, \frac{\partial u_{\epsilon}}{\partial \epsilon} \vert_{\epsilon=0}\rangle \Big]
\end{align*}

We have 
\begin{align}
\bar{K}_{\epsilon} u_{\epsilon} &= \lambda_{\epsilon} u _{\epsilon} \\
(\bar{K}_{\epsilon} - \lambda_{\epsilon} I) u _{\epsilon} &= 0 \\
\Big ( \frac{\partial \bar{K}_{\epsilon}}{\partial \epsilon}\vert_{\epsilon=0} -\frac{\partial \lambda_{\epsilon}}{\partial  \epsilon}\vert_{\epsilon=0} I\Big ) u + (\bar{K} - \lambda I) \frac{\partial u  _{\epsilon} }{\partial \epsilon}\vert_{\epsilon=0} &= 0 \\
\frac{\partial u  _{\epsilon} }{\partial \epsilon} \vert_{\epsilon=0} &=  (\lambda I-\bar{K})^{+} \Big ( \frac{\partial \bar{K}_{\epsilon}}{\partial \epsilon}\vert_{\epsilon=0}  -\frac{\partial \lambda_{\epsilon}}{\partial  \epsilon}\vert_{\epsilon=0}  I\Big ) u
\end{align}

Besides,
\begin{align}
v^T_{\epsilon}\bar{K}_{\epsilon}  &=  v^T_{\epsilon} \lambda_{\epsilon} \\
v^T_{\epsilon}(\bar{K}_{\epsilon} - \lambda_{\epsilon} I)  &= 0 \\
v^T \Big ( \frac{\partial \bar{K}_{\epsilon}}{\partial \epsilon}\vert_{\epsilon=0} -\frac{\partial \lambda_{\epsilon}}{\partial  \epsilon}\vert_{\epsilon=0} I\Big )  + \frac{\partial v^T _{\epsilon} }{\partial \epsilon}\vert_{\epsilon=0} (\bar{K} - \lambda I)  &= 0 \\
\frac{\partial v^T_{\epsilon} }{\partial \epsilon} \vert_{\epsilon=0} &=   v^T\Big ( \frac{\partial \bar{K}_{\epsilon}}{\partial \epsilon}\vert_{\epsilon=0}  -\frac{\partial \lambda_{\epsilon}}{\partial  \epsilon}\vert_{\epsilon=0}  I\Big ) (\lambda I-\bar{K})^{+}
\end{align}


We denote 
\begin{align}
W_{1,\theta(t),ij} =\frac{I\{Z_t=i,Z^*_{t+1}=j\} - I\{Z_t=i\} P(Z^*_{t+1}=j|Z_t=i,\theta(t))}{P(Z_t=i)} 
\end{align}
and
\begin{align}
W_{2,\theta(t),ij} =\frac{I\{Z_t=i\}\Big(Y_j- E(Y_j|Z_t=i,,\theta(t))\Big)}{P(Z_t=i)} 
\end{align}
for $i,j = 1,2, \cdots, N$.
Then the influence function for $\frac{\partial K_{\theta(t),\epsilon}}{\partial \epsilon} \vert_{\epsilon=0}$ is 
\begin{align}
 W_{1,\theta(t)} + W_{2,\theta(t)}
\end{align}
And the influence function for $\frac{\partial \bar{K}_{\epsilon}}{\partial \epsilon}\vert_{\epsilon=0}$  is 
\begin{align}
E_{\theta(t)} (W_{1,\theta(t)}) + E_{\theta(t)} (W_{2,\theta(t)}).
\end{align}
We introduce the notations
\begin{align}
v^T_{1} = v^T(\lambda I - \bar{K})^+ \qquad  v^T_{2} = v^T K_{\theta(t)} (\lambda I - \bar{K})^+  \\
u_{1} =  (\lambda I - \bar{K})^+ u \qquad u_{2} =  (\lambda I - \bar{K})^+ K_{\theta(t)} u  \\
c_{\theta(t)} = v^TK_{\theta(t)}u \qquad \bar{c}= \frac{v^T\Big( E_{\theta(t)} (W_{1,\theta(t)}) + E_{\theta(t)}(W_{2,\theta(t)}) \Big)u}{\langle v,u\rangle}
\end{align}

Then we have
\begin{align}
\phi(Z_{t},Z^*_{t+1},Y_1,\cdots,Y_N,\theta(t)) = \frac{C \langle v,u\rangle  - c_{\theta(t)}D}{c_{\theta(t)} \langle v,u\rangle}
\end{align}
where
\begin{align*}
C &:=  \Big [ v^T \Big(E_{\theta(t)} (W_{1,\theta(t)}) + E_{\theta(t)} (W_{2,\theta(t)}) -\bar{c}\Big) u_{2} + v^T\Big( W_{1,\theta(t)} + W_{2,\theta(t)}\Big)u + \\ & \qquad v^T_{2} \Big(E_{\theta(t)} (W_{1,\theta(t)}) + E_{\theta(t)} (W_{2,\theta(t)})  -\bar{c}\Big) u\Big]  \\
D &:=   \Big [ v^T \Big(E_{\theta(t)} (W_{1,\theta(t)}) + E_{\theta(t)} (W_{2,\theta(t)})  -\bar{c}\Big)u_{1} + v^T_1 \Big(E_{\theta(t)} (W_{1,\theta(t)}) + E_{\theta(t)} (W_{2,\theta(t)})  -\bar{c}\Big) u]
\end{align*}



\end{document}